\documentclass[%
reprint,
superscriptaddress,
 amsmath,amssymb,
 pra,
]{revtex4-2}

\usepackage{graphicx}
\usepackage{dcolumn}
\usepackage{bm}
\usepackage{hyperref}
\usepackage{placeins}
\usepackage{physics}
\usepackage{babel}
\usepackage{xcolor}
\usepackage{dsfont}
\usepackage{amsthm}
\usepackage{braket}
\usepackage{optidef}
\usepackage{algorithm2e}
\usepackage[toc,page]{appendix}
\usepackage[caption=false]{subfig}

\makeatletter
\let\ORIbbl@fixname\bbl@fixname
\def\bbl@fixname#1{%
  \@ifundefined{languagealias@\expandafter\string#1}
    {\ORIbbl@fixname#1}
    {\edef\languagename{\@nameuse{languagealias@#1}}}%
}
\newcommand{\definelanguagealias}[2]{%
  \@namedef{languagealias@#1}{#2}%
}
\makeatother

\newcommand{\me}{m_\textrm{e}}

\begin{document}

\preprint{APS/123-QED}

\title{Spin qubit gates via phonon buses in electron nanowires}

\author{Dylan Lewis}
\affiliation{Blackett Laboratory, Imperial College London, London, SW7 2AZ, United Kingdom}
\affiliation{Department of Physics and Astronomy, University College London, London WC1E 6BT, United Kingdom}

\author{Roopayan Ghosh}
\affiliation{Department of Physics and Astronomy, University College London, London WC1E 6BT, United Kingdom}
\affiliation{School of Basic Sciences, Indian Institute of Technology, Bhubaneswar, 752050, India.}

\author{Sanjeev Kumar}
\affiliation{Department of Electronic and Electrical Engineering, University College London, London WC1E 7JE, United Kingdom}

\author{Michael Pepper}
\affiliation{Department of Electronic and Electrical Engineering, University College London, London WC1E 7JE, United Kingdom}
\affiliation{London Centre for Nanotechnology, University College London, London WC1H 0AH, United Kingdom}

\author{Charles Smith}
\affiliation{Cavendish Laboratory, Department of Physics, University of Cambridge, Cambridge CB3 0HE, UK}

\author{Karyn Le Hur}
\affiliation{CPHT, CNRS, École Polytechnique, Institut Polytechnique de Paris, 91120 Palaiseau, France}

\author{Sougato Bose}
\affiliation{Department of Physics and Astronomy, University College London, London WC1E 6BT, United Kingdom}


\begin{abstract}
Scalable architectures for quantum computing using semiconductor quantum dots require interactions between qubits beyond adjacent quantum dots. Here, we propose using nanowires of electrons to mediate the interaction between two quantum dots. Virtual phonons in the linear chain of electrons can mediate an interaction that gives rise to effective spin-spin coupling of the electrons in distant quantum dots. We find coupling strengths of more than $30$~MHz for experimentally realisable parameters in GaAs quantum dots. 
\end{abstract}

\maketitle


\paragraph*{Introduction.} In theory, quantum computing can solve many computational problems more efficiently than a classical computer~\cite{montanaro_quantum_2016}. However, numerous engineering bottlenecks remain before a large-scale fault-tolerant quantum computer is achieved~\cite{garciadearquerSemiconductorQuantumDots2021}. Building qubits from spins in semiconductor quantum dots is certainly a promising approach. Recently, there has been important progress with demonstrations of high fidelity single- and two-qubit gates, initialisation, and readout~\cite{hendrickx_four-qubit_2021, philips_universal_2022}. Semiconductor quantum dots have the advantage of utilising advanced fabrication techniques from the mature CMOS integrated circuit industry. 

Each quantum dot is about 100 nanometres in size, allowing millions of physical qubits per chip. This introduces a significant scaling challenge. Controlling each quantum dot requires electrical signals, and since the qubits are closely packed, the wiring density for naively assembling a two-dimensional array of quantum dots is infeasible. Several architectures have been proposed to address this concern~\cite{cai_silicon_2019, jnane_multicore_2022, boter_spiderweb_2022, patomaki_pipeline_2023}. The architectures use the coherent shuttling of electrons~\cite{baart_single-spin_2016, fujita_coherent_2017, mills_shuttling_2019, buonacorsi_simulated_2020, ginzel_spin_2020, seidler_conveyor-mode_2021} or shared control of gate voltages~\cite{borsoi_shared_2023}. Coherent quantum state transfer schemes could also be used to transfer charges or spins and reduce the density of the control lines~\cite{Bose2003, antonioSelfAssembledWignerCrystals2015, lewis_low-dissipation_2023}. 
\begin{figure}[h!]
    \centering
    \includegraphics[scale=0.76]{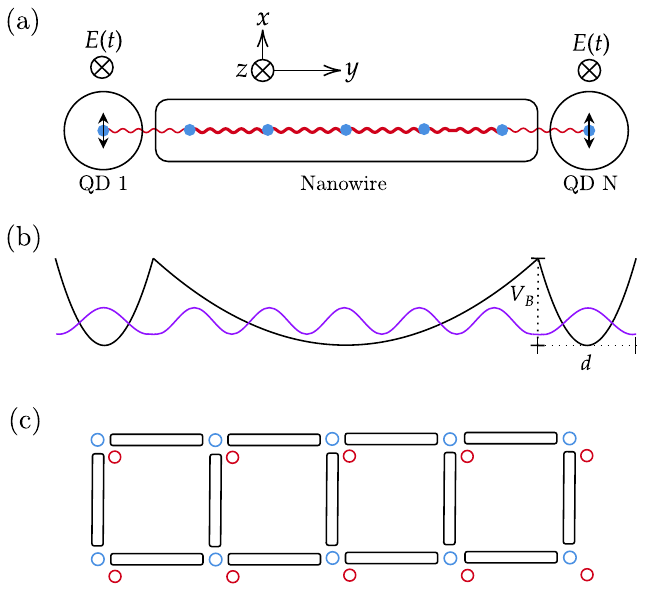}
    \caption{Illustrations of the electron bus. (a) Schematic of the electrons, labelling the end quantum dots as $1$ and $N$, corresponding to the number of electrons in the system. The electric field $E(t)$ in the $z$ direction is shown influencing quantum dots $1$ and $N$. The electric field is time-dependent in that it can be switched on, $E(t)=E_0$, and off, $E(t) = 0$, to switch on and off the Rashba effect. The axes are shown with $z$ as the out-of-plane axis. Coulomb repulsion between the electrons is depicted as springs and gives rise to the phonon modes. (b) Depicts the confining potential for the electrons and that they are in their ground state. The barrier height $V_B$ is determined by the confinement of the quantum dots and the nanowire. $d$ is the width of a quantum dot, which is typically around 100~nm in GaAs. (c) An example of a 2-dimensional planar architecture is given with electron nanowires (black rectangle boxes), and quantum dots, (blue and red). The red quantum dots can be used for measurement of the blue quantum dots.}
    \label{fig:electron_bus_diagram}
\end{figure}

Quantum dots (QDs) can be created by using electrodes in a GaAs/AlGaAs two-dimensional electron gas (2DEG), by using gates. This is rapidly emerging as a leading platform for spin-based quantum computing, offering long spin coherence times, precise electrostatic control, and compatibility with established semiconductor fabrication processes. However, the central challenge in scaling these systems is to implement high-fidelity two-qubit gates between spins separated by more than a few hundred nanometres, where direct exchange coupling becomes negligible, and thus also preventing crosstalk. Inspired by trapped ion experiments and a number of other proposals for quantum simulators~\cite{orthDissipativeQuantumIsing2008, orthDynamicsSynchronizationQuantum2010, wallBosonmediatedQuantumSpin2017, lewisIonTrapLongrange2023, bernhardtMajoranaFermionsQuantum2024}, including a recent proposal to trap electrons in a Paul trap~\cite{huangNumericalInvestigationsElectron2025}, we explore the possibility of using trapped electrons to mediate interactions between quantum dots. The interaction is mediated through a suspended electron nanowire that is electrostatically defined within the same 2DEG as the quantum dots. Such a nanowire can be formed by depleting the surrounding electron gas using patterned gate electrodes, confining electrons to a quasi-one-dimensional channel that supports quantised vibrational (phonon-like) modes.

\paragraph*{Linear electron chains.} 
The confining potentials of quantum dots can be modelled as harmonic~\cite{wensauer_laterally_2000, helle_two-electron_2005, li_exchange_2010, yang_generic_2011}. The potential of the longer one-dimensional nanowire is modelled as a quantum dot with significantly decreased confinement along one axis. The 2DEG can then be sufficiently depleted such that a single electron lies in each of first and last quantum dots; see Fig.~\ref{fig:electron_bus_diagram} for an illustration. A linear chain of electrons forms a nanowire in between the quantum dots. We model the nanowire as a longer depleted region of electrons -- essentially, an extended quantum dot with multiple electrons confined as a linear crystal. The electrons repel each other and this gives rise to the collective modes of motion -- the phonons. If the confinement of the nanowire is increased it changes the phonon modes and the bus becomes off-resonant, which gives a method for switching on and off the phonon bus. The potentials of the quantum dots and the wire are given by 
\begin{multline}
    V_{\textrm{trap}}(\bm{\omega}, \bm{r}_0) = \frac{1}{2} \sum_{j=1}^{N} m^* \big(\omega_{x}^2 (x_j-x_0)^2 \\ + \omega_{y}^2 (y_j-y_0)^2 + \omega_{z}^2 (z_j-z_0)^2 \big),
\end{multline}
where $\bm{\omega} = (\omega_{x}, \omega_{y}, \omega_{z})$ is the trapping frequency along the three spatial dimensions, $\bm{r}_j = (x_j, y_j, z_j)$ is the position of electron $j$ with trap centre $\bm{r}_0 = (x_0, y_0, z_0)$, and $m^*$ is the effective mass of the electron. In the following and throughout, we use $\hbar = 1$ and energy is given in frequency units rad/s. Typical parameters in GaAs quantum dots are an effective mass of $m^* = 0.067\me$, relative permittivity of $\epsilon_r = 12.9$, and a dot ground state energy of $\omega_\textrm{GS} \approx 6\times 10^{11}~\mathrm{rad/s}$, which gives a reasonably large dot of $300$~nm~\cite{reimann_electronic_2002}. The trapping frequency in the $z$ direction, $\omega_z$, is sufficient to confine the electron to a 2DEG. There are three separate trap centres: quantum dot 1, the electron nanowire, and quantum dot $N$. These have trap frequencies $\bm{\omega}_\textrm{a}$, $\bm{\omega}_\textrm{w}$, and $\bm{\omega}_\textrm{b}$, and trap centres $\bm{r}_\textrm{a}$, $\bm{r}_{\textrm{w}}$, and $\bm{r}_\textrm{b}$ respectively. 

The $x$ and $y$ trapping frequencies of the quantum dots are equal $\omega_{\textrm{a},x} = \omega_{\textrm{b},x}$ ($=\omega_{\textrm{a},y} = \omega_{\textrm{b},y}$), which is equal to the frequency $\omega_\textrm{GS}$ in $x$ and $y$. The nanowire has a reduced $y$ trapping frequency along its axial mode. For a wire of 6 electrons, the $y$ trapping frequency is $\omega_{\textrm{w},y} \approx \omega_\textrm{GS} / 10$. The nanowire has the same $x$  trapping frequency as the quantum dots. The centre of the nanowire trap is at the origin and the quantum dots are at $\bm{r}_\textrm{a} = (0, -1000,0)~\textrm{nm}$ and $\bm{r}_\textrm{b} = (0, 1000,0)~\textrm{nm}$.

\begin{figure*}
    \centering
    \includegraphics[scale=0.385]{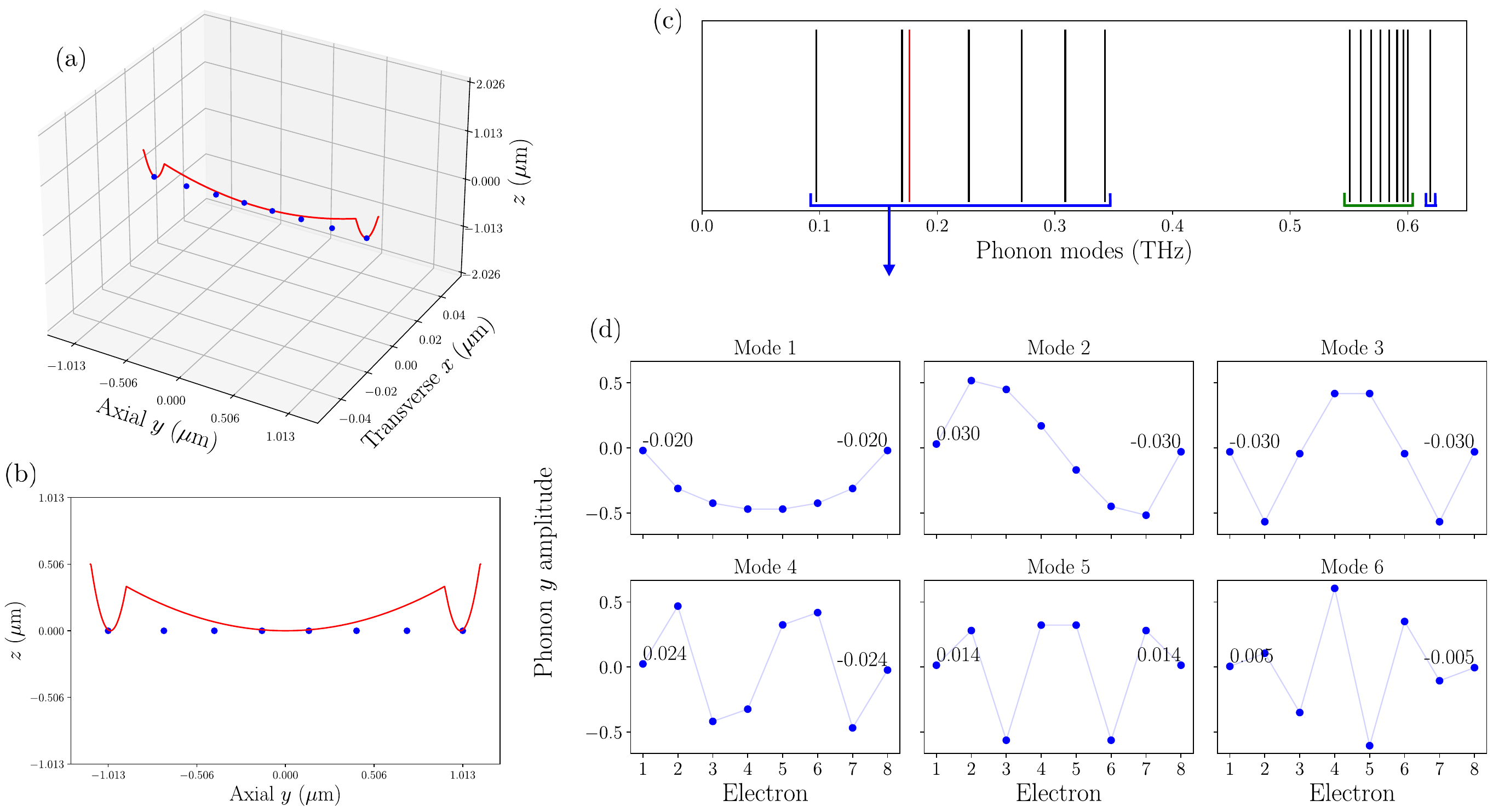}
    \caption{(a) Shows the positions of 8 electrons that minimise the confining potential of the quantum dots and nanowire (shown rescaled in red) for $\omega_{\textrm{w},y} = \omega_\textrm{GS} / 10$ with $\omega_\textrm{GS} = 0.6\times10^{11}~\textrm{rad/s}$, with nanowire trap at the origin and dot centres at $\bm{r}_\textrm{a} = (0, -1000,0)~\textrm{nm}$ and $\bm{r}_\textrm{b} = (0, 1000,0)~\textrm{nm}$. (b) Shows the $y$-$z$ projection of the electron positions and the rescaled confining potential. (c) Plots the resulting phonons, with the axial phonon modes (underlined in blue) and the transverse phonon modes (underlined in green). The red line is at the frequency of the Zeeman splitting, $\omega_0 = 1.765 \times10^{11}~\textrm{rad/s}$, induced by a magnetic field along $z$ with strength $5.151~\textrm{T}$, with $g=0.39$ and $m^*=0.067 m_\textrm{e}$. (d) Gives the corresponding phonon modes of the 8 electrons for the first 6 of the axial modes, mode 2 and 3 show the strongest and approximately equal contribution of the first and last electrons -- we chose mode 2 because it has a lower frequency.}
    \label{fig:phonon_modes_with_positions}
\end{figure*}

Since the electrons are cold, they simply assume the configuration that minimises the energy configuration of  
\begin{multline}
    V = \min\left\{ V_\textrm{trap}(\bm{\omega}_\textrm{a}, \bm{r}_\textrm{a}), V_\textrm{trap}(\bm{\omega}_\textrm{w}, \bm{r}_\textrm{w}), V_\textrm{trap}(\bm{\omega}_\textrm{b}, \bm{r}_\textrm{b})\right\} \\ + \frac{1}{2} \sum_{j=1}^N \sum_{i \ne j} \frac{e^2}{4\pi \epsilon_r \epsilon_0 |\bm{r}_i - \bm{r}_j |}, 
\end{multline}
where the overall trap potential is the minimum of the confinement potentials of the dots and nanowire, see Fig.~\ref{fig:electron_bus_diagram}(b).
Small vibrations around the equilibrium positions of the linear chain are quantised as excitations of phonon modes, $b_{y,i m}$, where phonon mode $m$ has frequency $\omega_m$, $i$ labels the electron, and an axial $y$ displacement is taken, but the same applies for any direction. The phonon modes are therefore calculated as the eigenvectors of the Hessian of $V$, which is the matrix of second-order partial derivatives of $V$ with respect to each of the three axes for each electron. The chain of the electrons is along the $y$ axis. A small displacement axially along $y$ from the equilibrium position of electron $i$ in the chain is given by 
\begin{equation}
    \label{eq:x_as_an_operator}
    y_i = \sum_m b_{y,i m} \hat{y}_m,
\end{equation}
where $\hat{y}_m$ is the operator of phonon mode $m$, given by 
\begin{equation}
    \label{eq:x_operator_in_phonons}
    \hat{y}_m = \xi^{(0)}_m (a^\dagger_m + a_m).
\end{equation}
The zero-mode spatial spread of phonon mode $m$ is $\xi^{(0)}_m = \sqrt{\frac{1}{2\me \omega_m}}$. The phonon modes $b_{i m}$ and phonon mode frequencies $\omega_m$ can be calculated as the eigenvectors of the solutions for small vibrations around the equilibrium positions of the linear chain~\cite{fishman_structural_2008}. In this way, the phonon modes are the eigenvectors of the Hessian matrix, with the phonon frequencies being the eigenvalues. Similarly, the momentum of electron $i$ is
\begin{align}
    p_i = \sum_{m} b_{y,im}\hat{p}_m,
\end{align}
where $\hat{p}_m$ is the momentum of phonon mode $m$,
\begin{align}
    \hat{p}_m = i \chi^{(0)}_m(a^{\dagger}_m - a_m),
\end{align}
with the zero-mode momentum spread of phonon mode $m$ given by $\chi_m^{(0)} = \sqrt{\frac{ \me \omega_m }{2}}$.  Results of the numerical minimisation of the confining potentials for GaAs quantum dots with the phonon mode spectrum is shown in Fig.~\ref{fig:phonon_modes_with_positions}. The code for the numerical simulation is given in Ref.~\cite{our_data}.

An external magnetic field causes a $z$-axis energy separation between spin up and spin down due to Zeeman splitting. A qubit can be encoded in the spin of the electron in the quantum dot with the up and down providing the two basis states. The energy level splitting induced by the magnetic field is $\omega_0 = 1.765\times10^{11}~\textrm{rad/s}$, requiring a magnetic field with strength $5.151~\textrm{T}$, which is readily available in experiments~\cite{pettaCoherentManipulationCoupled2005, hansonZeemanEnergySpin2003}, with a typical $g$-factor of $g = 0.39$~\cite{pioro-ladriereElectricallyDrivenSingleelectron2008}. 
For universal quantum computation, we require single-qubit gates and a two-qubit entangling gate. The simplest single-qubit gates we require is the ability to implement $R_x(\theta)$ and $R_y(\theta)$ rotations, which are rotations about the $x$-axis and $y$-axis of the Bloch sphere by the angle $\theta$. These rotation gates allow transitions between the qubit basis states and any superposition of both qubit states. The generators of the rotations are the $\sigma^x$ and $\sigma^y$ Hamiltonian terms. Changing the applied time corresponds to changing the angle $\theta$ of the rotations. The desired Hamiltonian terms can be induced directly with strong magnetic fields along the $x$ and $y$ axes. However, engineering strong localised magnetic fields in semiconductor devices that can be controlled precisely is experimentally difficult~\cite{jardineIntegratingMicromagnetsHybrid2021}. On the other hand, electric fields can straightforwardly be generated and controlled in semiconductor devices through gate voltages. As an alternative to generating local magnet fields directly through micromagnets, see App.~\ref{sec:micromagnetcs}, the idea here is to generate effective magnetic fields through the control of electric fields via spin-orbit coupling. While electron spins do not couple to electric fields directly, they couple indirectly through magnetic field gradients~\cite{ zhangControllingSyntheticSpinOrbit2021} or spin-orbit interactions~\cite{nowackCoherentControlSingle2007}. A spin-orbit interaction is the coupling of the electron motional degrees of freedom to its spin state. This is the approach we take here.

The intrinsic spin-orbit interaction in GaAs semiconductors is due to the Rashba and Dresselhaus effects,
\begin{align}
    H_{\textrm{SO}} = \alpha(p_x \sigma^y - p_y \sigma^x) + \beta(-p_x \sigma^x + p_y \sigma^y),
\end{align}
where $\alpha = -\frac{g \mu_\textrm{B} E_0}{2 m c^2}$ is the Rashba coefficient, $E_0$ is an applied time-independent electric field, $m$ is the electron mass, $\mu_\textrm{B}$ is the Bohr magneton, $\beta$ is the Dresselhaus coefficient, and $p_x$ ($p_y$) is the momentum in the $x$ ($y$) direction. Overall, the spin-orbit interaction can be considered an effective magnetic field. The values of $\alpha$ and $\beta$ are experimentally determined. In particular, $\alpha$ can be increased by strengthening the electric fields that give rise to an inversion asymmetry and are far larger than $\beta$, such that we can neglect $\beta$. There is also a symmetry between the Rashba effect and the Dresselhaus effect. The Dresselhaus effect ultimately results in a similar interaction form as the Rashba effect and would contribute to the total interaction strength. In Ref.~\cite{nowack_coherent_2007}, both Rashba and Dresselhaus were considered together in a single-electron case and can be expressed as an effective magnetic field.

\paragraph*{Single-qubit gates.}
A single-qubit gate using spin-orbit coupling from the Rashba and Dresselhaus terms was shown experimentally and described in Ref.~\cite{nowack_coherent_2007}. With a particular transformation, the Rashba and Dresselhaus terms can be viewed as a position-dependent effective magnetic field orthogonal to the applied external magnetic field~\cite{levitovDynamicalSpinelectricCoupling2002}. The external magnetic field is what gives rise to the Zeeman splitting. A time-dependent electric field is applied and induces adiabatic oscillations of the electron within the quantum dot. Due to the position dependence of the effective magnetic field, the electron feels an oscillating effective magnetic field. The direction of the effective magnetic field is perpendicular to the external magnetic field. Thus, Rabi oscillations can be observed as the spin state rotates. In this way, single qubit gates, $R_x(\theta)$ and $R_y(\theta)$ gates, can be implemented. Combinations of these rotations also give rise to $R_z(\theta)$ rotations.

\paragraph*{Two-qubit gates.}
We perform gates in the dispersive regime, which means that the Zeeman splitting $\omega_0$ is sufficiently detuned from the axial phonon mode frequency of interest, which is the \emph{second mode} with frequency $\omega_{y,2} = 1.70006\times10^{11}~\textrm{rad/s}$. The dispersive regime requires a small first-order Dyson series term such that phonons are only virtually required to mediate a spin-spin interaction between the two end quantum dots, see App.~\ref{sec:rashba_hamiltonian}. The detuning is $\Delta_2 = \omega_0 - \omega_{y,2} = 0.06459 \times10^{11}~\textrm{rad/s}$ and the coupling $g_{y, 1 2} = \alpha b_{y,1 2} \chi^{(0)}_{y,2} =  0.006324\times10^{11}~\textrm{rad/s}$ where the subscript $1$ label refers to electron 1, subscript $2$ label refers to phonon mode 2, and the Rashba coefficient is $\alpha = 1 \times 10^{-11}~$eVm~\cite{takaseHighlyGatetuneableRashba2017}. This gives a first-order Dyson term magnitude of $g_{y,12}/\Delta_{2} = 0.097$, which is below a typical threshold for the dispersive regime of 0.1. The resonant regime for $\Delta_2 \rightarrow 2$ generates real phonons and is discussed in App.~\ref{sec:resonance}. In the dispersive regime, the amplitude of real phonons generated would be small. Due to detuning, only one phonon mode has a significant coupling (phonon mode 2). The small amplitude of the phonon number oscillates and disentangles periodically at a high frequency~\cite{lewis_ion_2022}. In Fig.~\ref{fig:phonon_modes_with_positions}, we show the electrons in the quantum dots and nanowire confining potentials, and we plot the phonon modes with the Zeeman splitting frequency. Fig.~\ref{fig:phonon_modes_with_positions} shows why phonon mode 2 was chosen. Both phonon modes 2 and 3 have approximately equal contributions to the first and last electron displacements. However, phonon mode 2, $\omega_{y,2}$, is a lower frequency, which reduces the required magnetic field for the Zeeman splitting. The main contribution to the XY spin-spin interaction is due to the second phonon mode
\begin{align}
    |J_{1N}| &\approx \frac{g_{y,12} g_{y,82} \omega_{y,2}}{\omega_0^2 - \omega_{y,2}^2} \\
    &= \frac{g_{y,12}^2 \omega_{y,2}}{\omega_0^2 - \omega_{y,2}^2} = 30.2~\textrm{MHz}, \label{eq:8_electron_coupling}
\end{align}
where, due to the symmetry of the modes, $g_{y,12} = g_{y,82}$.
The full computation, considering all the phonon mode contributions, including the dispersive transverse $x$ phonon modes, gives an interaction strength of $J_{1N} = 33.9~\textrm{MHz}$. The full derivation of the Hamiltonian is given in App.~\ref{sec:rashba_hamiltonian}. The code for computing these results is given in Ref.~\cite{our_data}. 

\begin{figure}
    \centering
    \includegraphics[width=\linewidth]{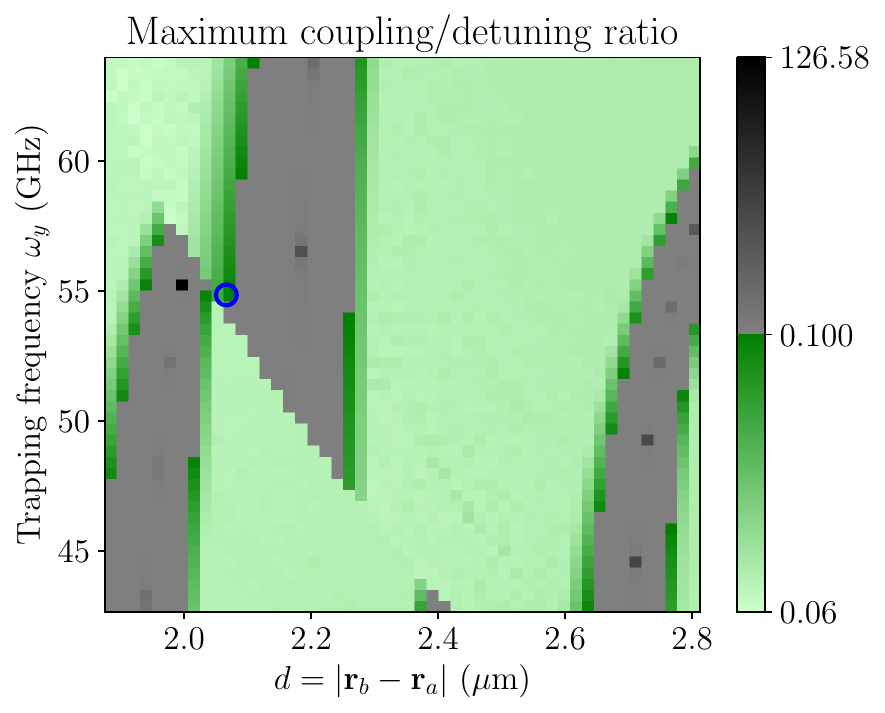}
    \caption{Ratio of maximum coupling to detuning by changing the axial trapping frequency $\omega_y$ and distance between the end quantum dots $d = |\bm{r}_b - \bm{r}_a|$ for $N=10$ electrons. The colour indicates the maximum coupling over detuning ratio value, which for phonon mode 2 is $g_{y,12}/\Delta_2$. Where the detuning $\Delta_2 = \omega_0 - \omega_{y,2}$, depends on the trapping frequency via the mode frequency $\omega_{y,2}$, and $\omega_0$ is kept constant. The green indicates a ratio less than 0.1, which defines the dispersive regime which gives our effective XY model coupling $J_{1N}$, see App.~\ref{sec:effective_hamiltonian}. The grey region is where the dispersive regime does not apply. The blue circle marks the maximum coupling $J_{1N}$ within the dispersive regime, where $g_{y,12}/\Delta_2 < 0.1$. The blue circle is for trapping frequency $\omega_y = 54.68~\mathrm{GHz}$ and distance between quantum dots $d = 2.066~\mathrm{\mu m}$, giving a maximum coupling strength $J_{1N} = 50.80~\mathrm{MHz}$.}
    \label{fig:parameter_sweep_n_10}
\end{figure}

\begin{figure}
    \centering
    \includegraphics[width=1\linewidth]{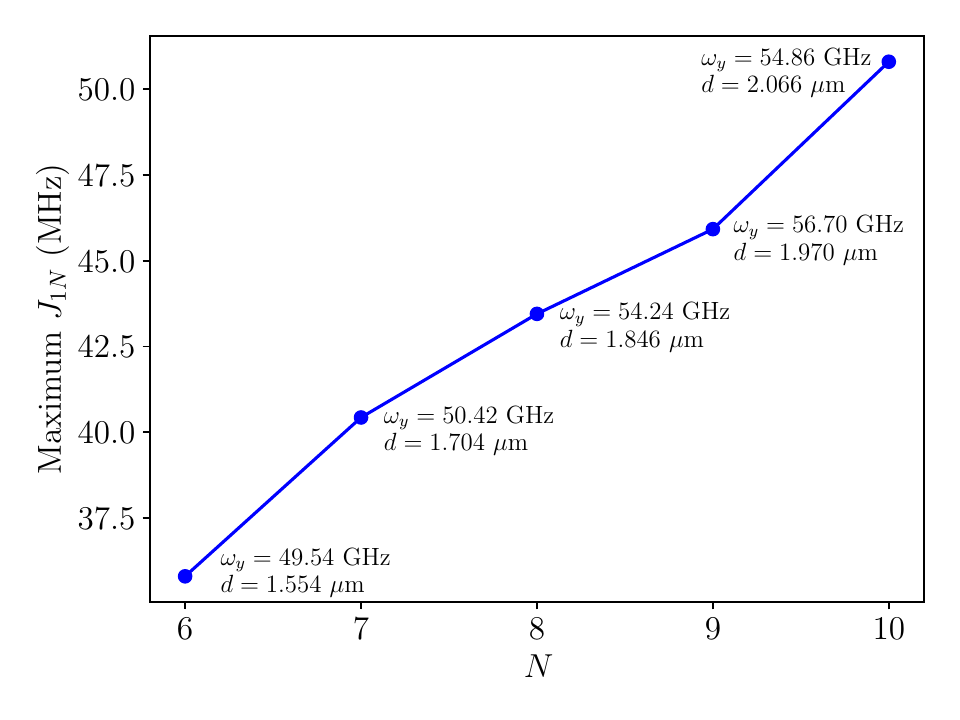}
    \caption{Maximum coupling strength $J_{1N}$ is plotted against the number of electrons $N$. At each point the trapping frequency $\omega_y$ and distance between end quantum dots $d = |\bm{r}_b - \bm{r}_a|$ is given. The maximum coupling value for $N=10$ is found using the parameter search in Fig.~\ref{fig:parameter_sweep_n_10}. In App.~\ref{sec:parameter_optimisation}, we show the parameters searches for the other values of $N$.}
    \label{fig:max_J_vs_N_electrons}
\end{figure}

The preceding results for Eq.~\eqref{eq:8_electron_coupling} and Fig.~\ref{fig:phonon_modes_with_positions} are for dots centred at $\bm{r}_\textrm{a} = (0, -1000,0)~\textrm{nm}$ and $\bm{r}_\textrm{b} = (0, 1000,0)~\textrm{nm}$ with 6 electrons in the wire and 1 electron in each end quantum dot, so $N=8$. However, if we allow the dot centres and axial nanowire trap frequency to be tuned, it is possible to find slightly higher coupling strengths. We can apply an optimisation procedure to nanowires with different numbers of electrons to find an approximate scaling of the coupling strength with nanowire length. We note that we have kept the Zeeman splitting $\omega_0$ and the trapping frequencies of the end quantum dots constant -- maintaining an approximate dot size of 300~nm. An example optimisation result is given in Fig.~\ref{fig:parameter_sweep_n_10} for 10 electrons. The maximum coupling strengths for a range of electron numbers are given in Fig.~\ref{fig:max_J_vs_N_electrons}. The optimisation results for 6,7,8, and 9 electrons are shown in App.~\ref{sec:parameter_optimisation}. Counterintuitively, the results show that as the chain increases in length the coupling actually gets stronger, up to over $50~$MHz for 10 electrons. A small chain with strong coupling requires a larger magnetic field (larger than is possible) because the lowest frequency modes are at a higher frequency than when additional electrons are added (and the centre nanowire is longer). The centre trap can therefore squeeze additional electrons together raising the frequency, while $\omega_0$ is kept constant, providing a stronger coupling. We find optimal coupling values that increase up to 10 electrons. This trend will not continue indefinitely, at some point the number of electrons is too great to be confined further with the axial trapping $\omega_y$ and whilst remaining a linear array of electrons. At this point, the coupling strength $J_{1N}$ will start to decrease with $N$.

\paragraph*{Lattice phonon noise.} 
In our GaAs quantum dot architecture, lattice phonon noise from deformation-potential coupling cannot exceed the level already present in established single-qubit gates. The dominant contribution from lattice vibrations arises from the first-order term in the Dyson series, which sets the leading-order electron–phonon interaction. In GaAs, the heavy Ga and As ions greatly suppress lattice displacement amplitudes for a given phonon energy, thereby reducing the deformation-potential coupling strength. In addition, the confined electrons used in our phonon-bus gate are not rigidly tied to the lattice, so the relevant electronic excitation frequencies are generally mismatched with the spectrum of bulk lattice phonons, further limiting their effect. Importantly, single-qubit gates in GaAs quantum dots which are already experimentally realized demonstrate fidelities which already include the full impact of lattice phonon noise. Since our phonon-bus–mediated two-qubit gate does not enhance this coupling pathway, the noise contribution from lattice phonons will necessarily remain below, or at most equal to, the level tolerated in the single-qubit gates.

\paragraph*{Discussion.}
Larger quantum dots (\(\sim 300~\mathrm{nm}\), compared to the more typical \(\sim 100~\mathrm{nm}\)) offer advantages for our proposal. They lower the magnetic field required to \(5.151~\mathrm{T}\) in order to obtain an $\omega_0$ resonant with a phonon mode, which in turn reduces the electron energies, moving them farther from the \(1\)–\(8~\mathrm{THz}\) range of lattice phonons~\cite{PhysRev.132.2410}, a common noise source. Our work is motivated by the challenge of scaling quantum dot architectures, where long-range coupling is often envisioned via microwave resonators. For qubits with \(\omega_0 = 1.765\times 10^{11}~\mathrm{rad/s}\), however, a half-wavelength resonator would be \(1.49~\mathrm{mm}\), about 5000 times the width of a quantum dot, making this approach impractical for dense integration. Hence our novel method offers a tremendous boost of scalability for quantum dot systems while offering intermediate speed gate operations. 

We have used the Rashba effect for generating the required spin-orbit coupling. It is also possible to use micromagnets to generate inhomogeneous magnetic fields that give rise to synthetic spin-orbit coupling. In App.~\ref{sec:micromagnetcs}, we derive the effective interaction for an electron phonon bus using micromagnets. The micromagnets method gives an Ising entangling interaction rather than the XY interaction of the method using the Rashba effect. However, since micromagnets pose a complex engineering challenge, we focus on the spin-orbit coupling using the Rashba effect here.

\paragraph*{Conclusion.} 
We have proposed a scalable method for long-range spin-spin interactions in quantum dots, mediated via phonon modes in electron nanowires. Electron nanowires are effectively linear arrays of electrons. Quantum dots can be connected via a nanowire. Virtual excitations of the collective motional modes of the electrons in the nanowire and quantum dots can be used to generate effective spin-spin interactions between the quantum dots. We show this with nanowires of between 4 and 8 electrons. In Fig.~\ref{fig:electron_bus_diagram} we have proposed a simple 2-dimensional array of quantum dots. In this schematic, there are two types of quantum dots. The quantum dots in red are not connected via the nanowires but are coupled to the quantum dots in blue via exchange interaction. Whereas the quantum dots in blue also interact with other quantum dots in blue via the nanowires in a 2-dimensional square lattice arrangement.

In theory, both the XY model generated by the Rashba effect and electric fields, as in App.~\ref{sec:rashba_hamiltonian}, and the Ising model with micromagnets, as in App.~\ref{sec:micromagnetcs}, could be applied simultaneously. In this case, anisotropic Heisenberg models would be possible that allow various topological phases~\cite{bernhardtMajoranaFermionsQuantum2024,dvirRealizationMinimalKitaev2023}.

For experimentally realistic parameters in GaAs quantum dots, we demonstrate coupling strengths of more than 30~MHz, which are sufficient for fast two-qubit gates. The nanowire gives an interaction distance between the quantum dots distance of around 2~$\mathrm{\mu m}$. The phonon bus scheme relies on electrostatic control and intrinsic spin-orbit coupling in quantum dots. This avoids the need for large microwave resonators, and allows the problem of the complex wiring bottleneck of quantum dots to be reduced. Overall, we show that electron nanowires, being used as phonon buses, could provide a method for fast intermediate-range entangling gates in scalable quantum dot architectures with minimal additional engineering requirements.

\section{Acknowledgments}
DL acknowledges support from the EPSRC Centre for Doctoral Training in Delivering Quantum Technologies, grant ref.~EP/S021582/1. SB, DL, MP and CGS acknowledge support from UK Research and Innovation (UKRI) Grant No. EP/R029075/1. SB and CGS acknowledge EPSRC-SFI funded project EP/X039889/1. SK is grateful to  the United Kingdom Research and Innovation (UKRI), Future Leaders Fellowship (MR/S015728/1; MR/X006077/1) for financial support.

\onecolumngrid
\appendix
\section*{Appendix}
\section{\label{sec:rashba_hamiltonian}Spin-spin interaction through Rashba effect}
\subsection{\label{sec:derivation_interaction_hamiltonian_rashba}Interaction Hamiltonian}
The Rashba effect stems from a breaking of the inversion symmetry in a 2-dimensional plane. If there is an out of plane electric field of strength $E_0$, along the $z$ axis, we have a Rashba term in the Hamiltonian of
\begin{align}
    H_\textrm{R} = \alpha (p_x \sigma^y - p_y \sigma^x),
\end{align}
with 
\begin{align}
    \alpha = -\frac{g \mu_\textrm{B} E_0}{2 m c^2},
\end{align}
where $m$ is the electron mass. The Rashba Hamiltonian gives the spin-orbit coupling term, 
\begin{align}
    H_\textrm{SOC} = \alpha (p_{x,1} \sigma^y_1 - p_{y,1} \sigma^x_1) + \alpha (p_{x,N} \sigma^y_N - p_{y,N} \sigma^x_N).
\end{align}
The assumption is that the spin-orbit coupling is perturbative and that the phonon modes established without the coupling remain. The non-perturbative part of the Hamiltonian is the phonons, $H_\textrm{ph}$, and the spins, $H_\textrm{sp}$,  
\begin{align}
    H_0 = \sum_{m} \omega_{x,m} a^\dagger_{x,m} a_{x,m} + \sum_{m} \omega_{y,m} a^\dagger_{y,m} a_{y,m} + \frac{\omega_0}{2}(\sigma^z_1 + \sigma^z_N),
\end{align}
where the Zeeman term that introduces energy level separation for the spins is applied along the out of plane direction of the electrons. The overall Hamiltonian is
\begin{align}
    H = \sum_{m} \omega_{x,m} a^\dagger_{x,m} a_{x,m} + \sum_{m} \omega_{y,m} a^\dagger_{y,m} a_{y,m} + \frac{\omega_0}{2}(\sigma^z_1 + \sigma^z_N) + \alpha (p_{x,1} \sigma^y_1 - p_{y,1} \sigma^x_1) + \alpha (p_{x,N} \sigma^y_N - p_{y,N} \sigma^x_N).
\end{align}
We can make one of the directions in the end quantum dots highly confined such that $\omega_{x,m}$ is too great to be excited. 

Transforming to the rotating frame of the spin Hamiltonian gives
\begin{align}
    H_{I_1} &= H_\textrm{ph} + e^{i\frac{\omega_0}{2}(\sigma^z_1 + \sigma^z_N)t}\alpha(p_{x,1} \sigma^y_1 + p_{x,N} \sigma^y_N ) e^{-i\frac{\omega_0}{2}(\sigma^z_1 + \sigma^z_N)t} \nonumber \\
     & \quad\quad\quad\quad\quad - e^{i\frac{\omega_0}{2}(\sigma^z_1 + \sigma^z_N)t}\alpha(p_{y,1} \sigma^x_1 + p_{y,N} \sigma^x_N ) e^{-i\frac{\omega_0}{2}(\sigma^z_1 + \sigma^z_N)t}  \\
    &= H_\textrm{ph} + \alpha p_{x,1} (\cos(\omega_0 t)\sigma^y_1 + \sin(\omega_0 t)\sigma^x_1) + \alpha p_{x,N} (\cos(\omega_0 t)\sigma^y_N + \sin(\omega_0 t)\sigma^x_N) \nonumber \\
    & \quad\quad\quad\quad\quad\quad\quad - \alpha p_{y,1} (\cos(\omega_0 t)\sigma^x_1 - \sin(\omega_0 t)\sigma^y_1) + \alpha p_{y,N} (\cos(\omega_0 t)\sigma^y_N - \sin(\omega_0 t)\sigma^x_N) \\
    &= H_\textrm{ph} + \alpha\sum_{j=\{1,N\}} \left[ i p_{x,j} (e^{-i\omega_0 t} \sigma^+_j - e^{i\omega_0 t} \sigma^-_j ) +  p_{y,j} (e^{i\omega_0 t} \sigma^+_j + e^{-i\omega_0 t} \sigma^-_j ) \right]
\end{align}
where we have used the identities,
\begin{align}
    e^{i\frac{\omega_0}{2}\sigma^z t} \sigma^x e^{-i\frac{\omega_0}{2}\sigma^z t} &= \cos(\omega_0 t)\sigma^x - \sin(\omega_0 t)\sigma^y\\
    e^{i\frac{\omega_0}{2}\sigma^z t} \sigma^y e^{-i\frac{\omega_0}{2}\sigma^z t} &= \cos(\omega_0 t)\sigma^y  + \sin(\omega_0 t)\sigma^x\\
    e^{i\frac{\omega_0}{2}\sigma^z t} \sigma^z e^{-i\frac{\omega_0}{2}\sigma^z t} &= \sigma^z.
\end{align}
The momenta can be written as 
\begin{align}
    p_{x,j} &= i\sum_{m} b_{x,jm}\chi^{(0)}_{x,m}(a^{\dagger}_{x,m} - a_{x,m}),\\
    p_{y,j} &= i\sum_{m} b_{y,jm}\chi^{(0)}_{y,m}(a^{\dagger}_{y,m} - a_{y,m}).
\end{align}
Transforming to the frame of the phonons gives
\begin{multline}
    H_{I_2} = \alpha\sum_{j=\{1,N\},m} \Big\{ b_{x,jm}\chi^{(0)}_{x,m}(e^{i\omega_{x,m} t} a^\dagger_{x,m} -e^{-i\omega_{x,m} t} a_{x,m}) (e^{i\omega_0 t} \sigma^+_j - e^{-i\omega_0 t} \sigma^-_j )  \\ + i b_{y,jm}\chi^{(0)}_{y,m}(e^{i\omega_{y,m} t} a^\dagger_{y,m} - e^{-i\omega_{y,m} t} a_{y,m}) (e^{i\omega_0 t} \sigma^+_j + e^{-i\omega_0 t} \sigma^-_j ) \Big\},
\end{multline}
where we have used
\begin{align}
    e^{i \omega_{x,m} a^\dagger_{x,m} a_{x,m} t } a_{x,m}e^{-i \omega_{x,m} a^\dagger_{x,m} a_{x,m} t } &= e^{-i\omega_{x,m} t} a_{x,m},\\
    e^{i \omega_{x,m} a^\dagger_{x,m} a_{x,m} t } a^\dagger_{x,m} e^{-i \omega_{x,m} a^\dagger_{x,m} a_{x,m} t } &= e^{i\omega_{x,m} t} a^\dagger_{x,m},
\end{align}
with equivalent transformations for $y$.
The Hamiltonian can be written as 
\begin{align}
    H_I = \sum_{j=\{1,N\},m} \Big\{ g_{x,jm}(e^{i(\omega_{x,m}+\omega_0) t} a^\dagger_{x,m} \sigma^+_j - e^{i(\omega_{x,m}-\omega_0) t} a_{x,m}^\dagger \sigma_j^- + h.c.) \\ + g_{y,jm}(i e^{i(\omega_{y,m} + \omega_0) t} a^\dagger_{y,m} \sigma_j^+ + i e^{i(\omega_{y,m}-\omega_0) t} a^\dagger_{y,m} \sigma_j^- + h.c. ) \Big\},
\end{align}
where we have defined $g_{x,jm} = \alpha b_{x,jm}\chi_{x,m}^{(0)}$ and $g_{y,jm} = \alpha b_{y,jm}\chi_{y,m}^{(0)}$.
We define two Hamiltonian terms 
\begin{align}
    H_{x} = \sum_{j=\{1,N\},m} g_{x,jm} (e^{i(\omega_{x,m}+\omega_0) t} a^\dagger_{x,m} \sigma^+_j - e^{i(\omega_{x,m}-\omega_0) t} a_{x,m}^\dagger \sigma_j^- + h.c.),
\end{align}
and
\begin{align}
    \label{eq:h_y}
    H_{y} = \sum_{j=\{1,N\},m}  g_{y,jm} (i e^{i(\omega_{y,m} + \omega_0) t} a^\dagger_{y,m} \sigma_j^+ + i e^{i(\omega_{y,m}-\omega_0) t} a^\dagger_{y,m} \sigma_j^- + h.c. ),
\end{align}
such that $H_I = H_x + H_y$.

\subsection{\label{sec:effective_hamiltonian}Effective Hamiltonian}
The effective Hamiltonian is found using the Dyson series expansion of the interaction Hamiltonian. The Dyson series is a solution to the Schrödinger equation with a time-dependent Hamiltonian. The lowest order terms of the power series of the exponential can be written
\begin{equation}
    U(t) = \mathds{1} - i \int_0^t d\tau_1 H_I(\tau_1) - \int_0^t d\tau_1 \int_0^{\tau_1} d\tau_2 H_I(\tau_1) H_I(\tau_2) + \dots
\end{equation}
The first non-trivial term is $U_1(t) = - i \int_0^t d\tau_1 H_I(\tau_1)$. We define
\begin{align}
    U_{1,x}(t) &= -i \int_0^t d\tau_1 H_x(\tau_1), \\
    U_{1,y}(t) &= -i \int_0^t d\tau_1 H_y(\tau_1),
\end{align}
and $U_{1}(t) =  U_{1,x}(t) + U_{1,y}(t)$. We find
\begin{align}
    U_{1,x}(t) &= -i\sum_{j=\{1,N\},m} g_{x,jm} ( c_{x,m}(1,1;t) a^\dagger_{x,m} \sigma^+_j - c_{x,m}(1,0;t) a_{x,m}^\dagger \sigma_j^- + h.c.)
\end{align}
where
\begin{align}
     c_{x,m}(p,q;t) &= \int^{t}_0 d\tau_1 e^{i (f_p\omega_{x,m}+f_q\omega_0) \tau_1} \\
     &=  \frac{i\left(1-e^{i(f_p\omega_{x,m}+f_q\omega_0)t}\right)}{(f_p\omega_{x,m}+f_q\omega_0)}, \label{eq:c_definition}
\end{align}
and we have defined a simple function for convenience $f_k = (-1)^{k+1}$. It is clear that 
\begin{equation}
    c_{x,m}(p,q;t)^* = c_{x,m}(p+1,q+1;t).
\end{equation} 
An equivalent expression can be derived for the $y$ term,
\begin{align}
    U_{1,y}(t) &= -i \sum_{j=\{1,N\},m} g_{y,jm} (i c_{y,m}(1,1;t) a^\dagger_{y,m} \sigma^+_j + i c_{y,m}(1,0;t) a_{y,m}^\dagger \sigma_j^- + h.c.).
\end{align}
The second order term is 
\begin{align}
    U_{2}(t) &=  -\int_0^t d\tau_1 \int_0^{\tau_1} d\tau_2 \left[ H_x(\tau_1) H_x(\tau_2)  +  H_x(\tau_1) H_y(\tau_2)   + H_y(\tau_1) H_x(\tau_2) + H_y(\tau_1) H_y(\tau_2) \right]\\
    &=-i\int_0^t d\tau_1 \left[ H_x(\tau_1) U_{1,x}(\tau_1)  +  H_x(\tau_1) U_{1,y}(\tau_1)   + H_y(\tau_1) U_{1,x}(\tau_1) + H_y(\tau_1) U_{1,y}(\tau_1) \right].
\end{align}
We define the terms 
\begin{align}
    U_{2,xx} &= -i\int_0^t d\tau_1  H_x(\tau_1) U_{1,x}(\tau_1) \\
    U_{2,xy} &= -i\int_0^t d\tau_1  H_x(\tau_1) U_{1,y}(\tau_1) \\
    U_{2,yx} &= -i\int_0^t d\tau_1  H_y(\tau_1) U_{1,x}(\tau_1) \\
    U_{2,yy} &= -i\int_0^t d\tau_1  H_y(\tau_1) U_{1,y}(\tau_1),
\end{align}
such that $U_2 = U_{2,xx} + U_{2,xy} + U_{2,yx} + U_{2,yy}$. The last term is 
\begin{multline}
    U_{2,yy} = - \int_0^t \left(\sum_{j=\{1,N\},l} g_{y,kl} (ie^{i(\omega_{y,l}+\omega_0) t} a^\dagger_{y,l} \sigma^+_k + i e^{i(\omega_{y,l}-\omega_0) t} a_{y,l}^\dagger \sigma_k^- + h.c.) \right) \\ \times \left(\sum_{j=\{1,N\},m} g_{y,jm} (i c_{y,m}(1,1;t) a^\dagger_{y,m} \sigma^+_j + i c_{y,m}(1,0;t) a_{y,m}^\dagger \sigma_j^- + h.c.)\right),
\end{multline}
which gives
\begin{multline}
     U_{2,yy} = \sum_{j,k=\{1,N\}, m,l} g_{y,jm} g_{y,kl}\Bigg(d_{y,lm}(1,1,1,1;t)a^\dagger_{y,l}a^\dagger_{y,m} \sigma^+_k \sigma^+_j + d_{y,lm}(1,1,1,0;t)a^\dagger_{y,l}a^\dagger_{y,m} \sigma^+_k \sigma^-_j \\
     + d_{y,lm}(1,1,0,1;t)a^\dagger_{y,l}a_{y,m} \sigma^+_k \sigma^+_j + d_{y,lm}(1,1,0,0;t)a^\dagger_{y,l}a_{y,m} \sigma^+_k \sigma^-_j \\
     + d_{y,lm}(1,0,1,1;t)a^\dagger_{y,l}a^\dagger_{y,m} \sigma^-_k \sigma^+_j + d_{y,lm}(1,0,1,0;t)a^\dagger_{y,l}a^\dagger_{y,m} \sigma^-_k \sigma^-_j \\
     +d_{y,lm}(1,0,0,1;t)a^\dagger_{y,l}a_{y,m} \sigma^-_k \sigma^+_j + d_{y,lm}(1,0,0,0;t)a^\dagger_{y,l} a_{y,m} \sigma^-_k \sigma^-_j + h.c. \Bigg),
\end{multline}
where the integral terms have been defined
\begin{align}
    d_{y,lm}(r,s,p,q;t) &= \int_0^t d\tau_1 c_{y,m}(p,q;t) e^{i(f_ r\omega_{y,m} + f_ s \omega_0)\tau_1} \\
    &= \frac{1}{f_p\omega_{y,m} + f_q \omega_0} \left( \frac{1-e^{i(f_r \omega_{y,l} + f_p \omega_{y,m} + (f_q + f_s)\omega_0) t}}{f_r \omega_{y,m} + f_p \omega_{y,l} + (f_q + f_s)\omega_0} - \frac{1- e^{i(f_r\omega_{y,l} + f_s\omega_0)t}}{f_r\omega_{y,l} + f_s\omega_0} \right).
\end{align}
In the end, only the secular terms that scale with $t$ will be relevant for virtual excitations of the phonon mode -- when $\omega_0$ is detuned from the phonon modes. Thus, we consider $m = l$ and  $p+r=1$ and $q+s=1$ only, giving, for example,
\begin{align}
    d_{mm}(1,0,0,1;t) = \frac{i t}{\omega_{y,m}-\omega_0} + \frac{1-e^{-i(\omega_{y,m}-\omega_{0})t}}{(\omega_{y,m}-\omega_0)^2}. 
\end{align}
Furthermore, we use the rotating wave approximation because the terms with $(\omega_{x,m}+\omega_{0})^2$ in the denominator are suppressed. This reduces the number of terms that have a significant contribution
\begin{multline}
    \tilde{U}_{2,yy} \approx \sum_{j,k=\{1,N\}, m} g_{y,jm} g_{y,km}\Bigg(d_{y,mm}(1,0,0,1;t)a^\dagger_{y,m}a_{y,m} \sigma^-_k \sigma^+_j + d_{y,mm}(0,1,1,0;t) a_{y,m}a^\dagger_{y,m} \sigma^+_k \sigma^-_j \\
    + d_{y,mm}(1,1,0,0;t) a^\dagger_{y,m}a_{y,m} \sigma^+_k \sigma^-_j + d_{y,mm}(0,0,1,1;t) a_{y,m}a^\dagger_{y,m} \sigma^-_k \sigma^+_j  \Bigg).
\end{multline}
After, including only the secular terms, we find
\begin{multline}
    \tilde{U}_{2,yy} \approx \sum_{j,k=\{1,N\}, m} g_{y,jm} g_{y,km}\Bigg(\frac{it}{\omega_{y,m} - \omega_0}a^\dagger_{y,m}a_{y,m} \sigma^-_k \sigma^+_j - \frac{it}{\omega_{y,m} - \omega_0} a_{y,m}a^\dagger_{y,m} \sigma^+_k \sigma^-_j \\
    + \frac{it}{\omega_{y,m} + \omega_0} a^\dagger_{y,m}a_{y,m} \sigma^+_k \sigma^-_j - \frac{it}{\omega_{y,m} + \omega_0} a_{y,m}a^\dagger_{y,m} \sigma^-_k \sigma^+_j  \Bigg).
\end{multline}
We then use the bosonic commutation relation $[a_m, a_l^\dagger] = \delta_{ml}$,
\begin{multline}
    \tilde{U}_2(t) \approx i t \sum_{j,k=\{1,N\},m}   g_{y,jm} g_{y,km} \left(\frac{2\omega_{y,m}}{\omega_0^2 - \omega_{y,m}^2}(\sigma_k^+ \sigma_j^- + \sigma_k^- \sigma_j^+) \right) \\
   + i t \sum_{j=\{1,N\},m}  g_{y,jm}^2 \left(a^\dagger_{y,m} a_{y,m} \frac{2\omega_0}{\omega_0^2 - \omega_{y,m}^2}\left( \sigma_j^+ \sigma_j^- - \sigma_j^- \sigma_j^+ \right)+ \frac{\sigma_j^+ \sigma_j^-}{\omega_0 - \omega_{y,m}} - \frac{\sigma_j^- \sigma_j^+}{\omega_0 + \omega_{y,m}}\right),
\end{multline}
This expression can be simplified with $\sigma^+ \sigma^- = \frac{1}{2}\left(\sigma^z + \mathds{1}\right)$, $\sigma^- \sigma^+ = - \frac{1}{2}\left(\sigma^z - \mathds{1}\right)$, and $\sigma_j^+ \sigma_i^- + \sigma_j^- \sigma_i^+ = \frac{1}{2}\left(\sigma_j^x \sigma_i^x + \sigma_j^y \sigma_i^y \right)$ to give
\begin{equation}
    \tilde{U}_{2,yy}(t) \approx  i t \sum_{m}   \frac{g_{y,1 m}g_{y, N m} \omega_{y,m}}{(\omega_0^2 - \omega_{y,m}^2)} \left( \sigma_N^x \sigma_1^x + \sigma_N^y \sigma_1^y \right)
    + i t \sum_{j=\{1,N\},m}  g_{y,jm}^2 \left(\frac{\omega_0}{\omega_0^2 - \omega_{y,m}^2} \left(2\hat{n}_{y,m} + 1\right) \sigma_j^z - \frac{\omega_{y,m}}{\omega_0^2 - \omega_{y,m}^2} \mathds{1} \right),
\end{equation}
where $\hat{n}_{y,m} = a^\dagger_{y,m} a_{y,m}$ is the number operator.
We therefore have the effective Hamiltonian between the first and last quantum dots
\begin{equation}
    \label{eq:pure_xy_spin_hamiltonian_app}
    H_{\textrm{XY}} =  J_{1 N}  \left( \sigma_1^x \sigma_N^x + \sigma_1^y \sigma_N^y \right) + h_1 \sigma_1^z + h_N \sigma_N^z,
\end{equation}
where $J_{1 N}= \sum_{m} \frac{g_{y,1 m}g_{y, N m} \omega_{y,m}}{(\omega_{y,m}^2 - \omega_0^2)}$, $h_j = \sum_{m} \frac{g_{y,j m}^2 \omega_0 }{4(\omega_{y,m}^2 - \omega_0^2)} \left( 2n + 1 \right)$, $n$ approximates the initial phonon number, and the identity term has been dropped. If $h_1$ and $h_N$ are equal it will only add a global phase to the coupling. As previously defined $g_{y,jm} =\alpha b_{y,jm}\chi_{y,m}^{(0)}$. This Hamiltonian only considers the $U_{2,yy}$ term. However, if only one term of the Rashba effect is permitted due to increased confinement, this is the effective Hamiltonian.
\subsection{\label{sec:resonance}At resonance}
The dynamics at resonance for the longitudinal phonon mode 2 can be seen by taking the limit $\omega_{y,2} \rightarrow \omega_0$ in Eq.~\eqref{eq:h_y}, which gives the interaction Hamiltonian
\begin{multline}
    H_y = \sum_{j=\{1,N\}} g_{y,j 2} (i e^{2i\omega_0 t} a^\dagger_{y,2} \sigma_j^+ + i a^\dagger_{y,2} \sigma_j^- + h.c. ) \\ +  \sum_{j=\{1,N\},m\ne2}  g_{y,jm} (i e^{i(\omega_{y,m} + \omega_0) t} a^\dagger_{y,m} \sigma_j^+ + i e^{i(\omega_{y,m}-\omega_0) t} a^\dagger_{y,m} \sigma_j^- + h.c. ).
\end{multline}
Taking the rotating wave approximation gives
\begin{align}
    H_y = \sum_{j=\{1,N\}} g_{y,j 2} ( i a^\dagger_{y,2} \sigma_j^- - i a_{y,2} \sigma_j^+ ) +  \sum_{j=\{1,N\},m\ne2}  g_{y,jm} ( i e^{i(\omega_{y,m}-\omega_0) t} a^\dagger_{y,m} \sigma_j^- + h.c. ).
\end{align}
The first term is just the interaction term of the Jaynes-Cummings model at resonance for a single boson mode and two spins. Oscillations occur between excited electron spin states and phonons in mode 2. 
Both electron spin states are equally coupled to the phonon mode. 
At arbitrary times the spin and phonon mode are entangled. Stroboscopically, the phonon state disentangles and returns to its original state. The times at which the phonon mode disentangles and the corresponding evolution of the spin states depends on the initial state. 
Generally, the resonant regime is difficult in our case for qubits encoded as spins because real phonons are generated at intermediate times, and it does not display a general effective spin-spin interaction irrespective of initial state. 
\newpage
\section{\label{sec:parameter_optimisation}Parameter optimisation}
\begin{figure}[h]
    \centering
    \subfloat[N = 6]{\includegraphics[width=0.48\linewidth]{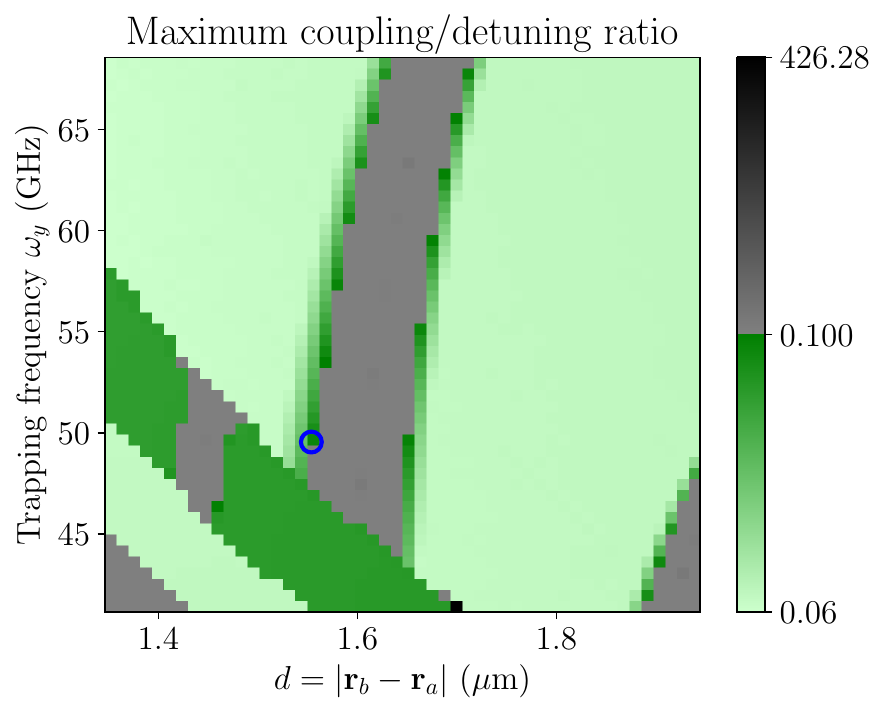}} \hspace{0.1cm}
    \subfloat[N = 7]{\includegraphics[width=0.48\linewidth]{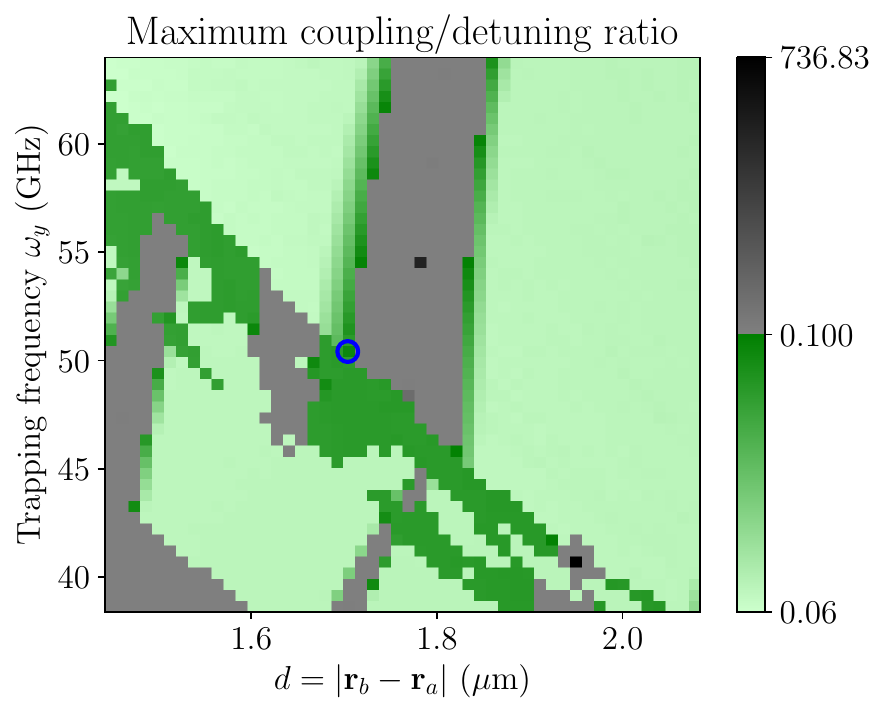}} \\
    \subfloat[N = 8]{\includegraphics[width=0.48\linewidth]{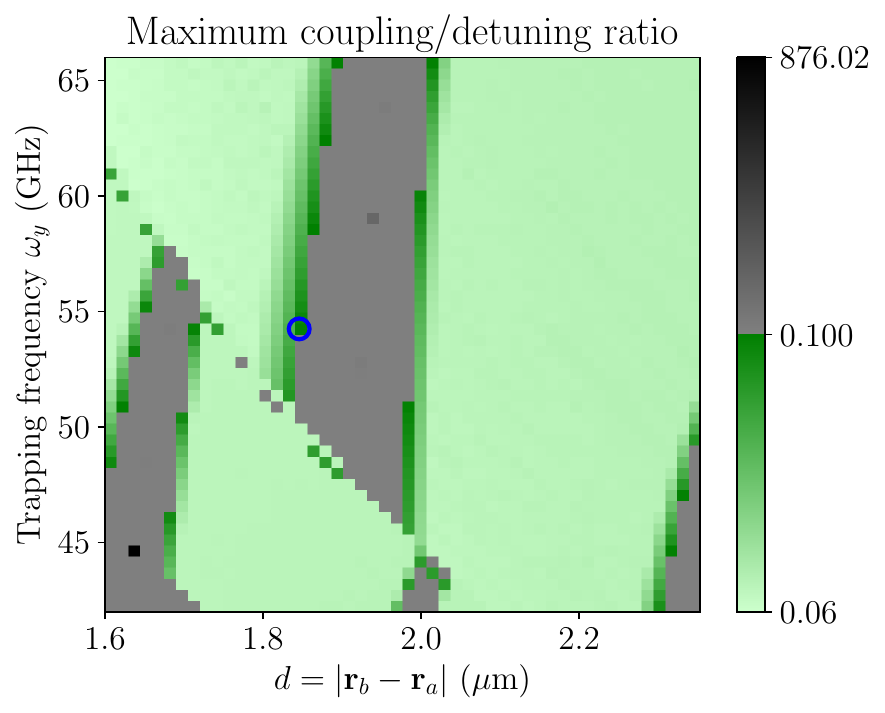}} \hspace{0.1cm}
    \subfloat[N = 9]{\includegraphics[width=0.48\linewidth]{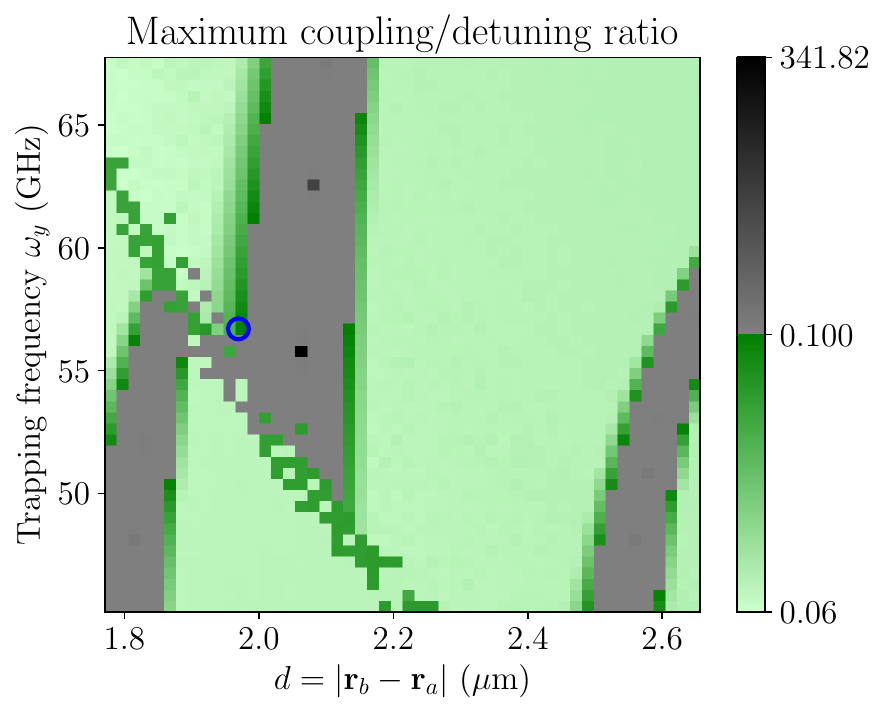}} 
    \caption{Ratio of maximum coupling to detuning by changing trapping frequency $\omega_y$ and distance between the end quantum dots $d = |\bm{r}_b - \bm{r}_a|$. The colour indicates the maximum coupling over detuning ratio value, which for phonon mode 2 is $g_{y,12}/\Delta_2$. Where the detuning $\Delta_2 = \omega_0 - \omega_{y,2}$, depends on the trapping frequency via the mode frequency $\omega_{y,2}$, and $\omega_0$ is kept constant. The green indicates a ratio less than 0.1, which defines the dispersive regime that gives the effective XY model coupling $J_{1N}$, see App.~\ref{sec:effective_hamiltonian}. The grey region is where the dispersive regime does not apply. The blue circle marks the maximum coupling $J_{1N}$ within the dispersive regime, where $g_{y,12}/\Delta_2 < 0.1$. The result of the blue circle gives the maximum $J_{1N}$ used in Fig.~\ref{fig:max_J_vs_N_electrons} in the main text.}
    \label{fig:parameter_optimisations}
\end{figure}

We assume a fixed magnetic field of $5.151~$T, which is readily achievable for Zeeman splitting. The two main parameters that can then be optimised for coupling strength are the trapping frequency $\omega_y$ and the distance between the end quantum dots $d = |\bm{r}_b - \bm{r}_a|$. In Fig.~\ref{fig:parameter_optimisations}, we show the parameter sweeps for $N=6,7,8,9$ electrons.
\FloatBarrier
\section{\label{sec:micromagnetcs}Spin-spin interaction through micromagnets}
\subsection{\label{sec:derivation_interaction_hamiltonian}Derivation of interaction Hamiltonian}
Micromagnets surrounding the end quantum dots produce homogeneous and inhomogeneous magnetic fields~\cite{huang_spin_2021}, $\bm{B}_j = \bm{\beta}_j + \bm{b}_j$ for site $j$, where $\bm{\beta}_j$ is the  homogeneous field and $\bm{b}_j=\bm{b}_j(\bm{r})$ is the inhomogeneous field.  The homogeneous field gives a Zeeman splitting that contributes to the qubit energy gap $\omega_0$. The inhomogeneous field gives rise to a synthetic spin-orbit coupling (SOC), to first order in position,
\begin{equation}
    H_{\textrm{SOC}} = \frac{1}{2}g\mu_B (\bm{\sigma}_1 \cdot\bm{\delta B}_1 + \bm{\sigma}_N \cdot\bm{\delta B}_N)
\end{equation}
with $\bm{\sigma}_j = (\sigma^x_j, \sigma^y_j, \sigma^z_j)$, $\bm{\delta B}_j = \frac{\partial \bm{B}_j}{\partial x}\big\vert_{x_j=0} x_j$ is the gradient of the inhomogeneous field times the displacement from equilibrium. Only magnetic fields acting on the quantum dots at the ends of the electron nanowire are considered. The magnetic field gradient is assumed to be in the $x$ coordinate axis. The assumption is that the spin-orbit coupling Hamiltonian is perturbative -- that is the phonon modes established without the spin-orbit coupling remain. 

We can transform to the rotating frame of the non-perturbative Hamiltonian $H_0 = H_{\textrm{ph}} + H_{\textrm{sp}}$. Overall the Hamiltonian is
\begin{align}
    H &= H_0 + H_\textrm{SOC} \\
    &= \sum_{m} \omega_m a^\dagger_m a_m + \frac{\omega_0}{2}\left( \sigma_1^z + \sigma_N^z \right) + \frac{1}{2}g\mu_B (\bm{\sigma}_1 \cdot\bm{\delta B}_1  + \bm{\sigma}_N \cdot\bm{\delta B}_N ).
\end{align}
First, we rotate to the frame of the spins,
\begin{align}
    H_{I_1} &= H_\textrm{ph} + e^{i\frac{\omega_0}{2}(\sigma^z_1 + \sigma^z_N)t}\frac{1}{2}g\mu_B (\bm{\sigma}_1 \cdot\bm{\delta B}_1 + \bm{\sigma}_N \cdot\bm{\delta B}_N ) e^{-i\frac{\omega_0}{2}(\sigma^z_1 + \sigma^z_N)t} \\
    &= H_\textrm{ph} + \frac{1}{2}g\mu_B (\bm{\sigma}_1^\prime(t) \cdot\bm{\delta B}_1 + \bm{\sigma}_N^\prime(t) \cdot\bm{\delta B}_N )
\end{align}
where we have used the identities,
\begin{align}
    e^{i\frac{\omega_0}{2}\sigma^z t} \sigma^x e^{-i\frac{\omega_0}{2}\sigma^z t} &= \cos(\omega_0 t)\sigma^x - \sin(\omega_0 t)\sigma^y\\
    e^{i\frac{\omega_0}{2}\sigma^z t} \sigma^y e^{-i\frac{\omega_0}{2}\sigma^z t} &= \cos(\omega_0 t)\sigma^y  + \sin(\omega_0 t)\sigma^x\\
    e^{i\frac{\omega_0}{2}\sigma^z t} \sigma^z e^{-i\frac{\omega_0}{2}\sigma^z t} &= \sigma^z,
\end{align}
to define the rotation transformation of the spins $\bm{\sigma}^\prime_j(t) =  U(t) \bm{\sigma}_j$ with
\begin{align}
    U = \begin{pmatrix}
        \cos(\omega_0 t) & -\sin(\omega_0 t)& 0\\
        \sin(\omega_0 t) & \cos(\omega_0 t) & 0\\
        0 & 0 & 1
    \end{pmatrix}.
\end{align}
We note that 
\begin{align}
    e^{i \sum_m \omega_m a_m^\dagger a_m t} x_j e^{-i \sum_m \omega_m a_m^\dagger a_m t} &= e^{i \sum_m \omega_m a_m^\dagger a_m t} \sum_{l} b_{j l} \hat{x}_l e^{-i \sum_m \omega_m a_m^\dagger a_m t}\\ 
    &= e^{i \sum_m \omega_m a_m^\dagger a_m t} \sum_{l} b_{j l} \xi_l^{(0)} (a^\dagger_l + a_l) e^{-i \sum_m \omega_m a_m^\dagger a_m t} \\
    &= \sum_{m} b_{j m} \xi_m^{(0)} \left(e^{i \omega_m t} a^\dagger_m + e^{- i \omega_m t} a_m \right),
\end{align}
using Eqs.~\eqref{eq:x_as_an_operator} and~\eqref{eq:x_operator_in_phonons}. Transforming to the rotating frame of the phonons thus gives
\begin{align}
    H_{I_2} &= \sum_{j=\{1,N\}}\sum_{m}  \mathcal{B}_{jm} \bm{\sigma}_j^\prime(t) \cdot \frac{\partial \bm{B}_j}{\partial x}\bigg\vert_{x_j=0} \left(e^{i \omega_m t} a^\dagger_m + e^{- i \omega_m t} a_m \right) 
\end{align}
where
\begin{align}
    \mathcal{B}_{jm} = \frac{1}{2}g \mu_B  b_{j m} \xi_m^{(0)}.
\end{align}
\subsection{\label{sec:effective_hamiltonian_rashba}Effective Hamiltonian}
The effective Hamiltonian is found using the Dyson series expansion of the interaction Hamiltonian. The Dyson series is a solution to the Schrödinger equation with a time-dependent Hamiltonian. The lowest order terms of the power series of the exponential can be written
\begin{equation}
    U(t) = \mathds{1} - i \int_0^t d\tau_1 H_I(\tau_1) - \int_0^t d\tau_1 \int_0^{\tau_1} d\tau_2 H_I(\tau_1) H_I(\tau_2) + \dots
\end{equation}
The first non-trivial term is $U_1(t) = - i \int_0^t d\tau_1 H_I(\tau_1)$. We find
\begin{align}
    U_{1,B}(t) &= - i \int_0^t d\tau_1 H_I(\tau_1) \\
    &= -i \sum_{j=\{1,N\}}\sum_{m}  \mathcal{B}_{jm} \bm{\sigma}_j^\prime(t) \cdot \frac{\partial \bm{B}_j}{\partial x}\bigg\vert_{x_j=0} \left( \alpha^*_B(t) a^\dagger_m + \alpha_B(t) a_m \right),
\end{align}
where we have defined the integral solution terms 
\begin{align}
    \alpha_B(t) &= \int_0^t d\tau_1 e^{-i \omega_m \tau_1} \\
    &= \frac{i(e^{-i \omega_m t} - 1)}{\omega_m},
\end{align}
with the function $f_k = (-1)^{k+1}$ to determine the sign. 

These solutions are oscillatory when $\omega_E$ is detuned from $\omega_m$. The second order Dyson series term gives
\begin{align}
    U_{2}(t) &= -i \int_0^t d\tau_1 H_B(\tau_1) U_{1}(\tau_1) \\
    &= -  \sum_{k,j=\{1,N\}}\sum_{l,m} \mathcal{B}_{kl} \mathcal{B}_{jm} \bm{\sigma}_k^\prime \cdot\bm{\delta B}_k   \bm{\sigma}_j^\prime \cdot\bm{\delta B}_j \int_0^t d\tau_1 \left(e^{i \omega_l \tau_1} a^\dagger_l + e^{- i \omega_l \tau_1} a_l \right) \left( \alpha^*_B(\tau_1) a^\dagger_m - \alpha_B(\tau_1) a_m \right),
\end{align}
firstly, we know that there will only be an spin-spin interaction mediated by a phonon at second order if the phonon modes are the same, so $m = l$, we therefore drop the terms with $m \ne l$ for the effective Hamiltonian,
\begin{align}
    \tilde{U}_{2}(t) &= - \sum_{k,j=\{1,N\}}\sum_{m} \mathcal{B}_{km} \mathcal{B}_{jm} \bm{\sigma}_k^\prime \cdot\bm{\delta B}_k   \bm{\sigma}_j^\prime \cdot\bm{\delta B}_j \left(\beta_{BB}(t) a_m a_m + \beta_{B\bar{B}}(t) a_m a_m^\dagger  + \beta_{B\bar{B}}^*(t) a_m a_m^\dagger +  \beta_{BB}^*(t) a_m^\dagger a_m^\dagger \right),
\end{align}
where we define second order integral solutions
\begin{align}
    \beta_{BB}(t) &= \int_0^t d\tau_1 e^{-i \omega_m\tau_1} \alpha_B(\tau_1) \\
    &= \frac{e^{-i\omega_m t}(1-\cos(\omega_m t))}{\omega_m^2},
\end{align}
and 
\begin{align}
    \beta_{B\bar{B}}(t) &= \int_0^t d\tau_1 e^{-i \omega_m\tau_1} \alpha_B^*(\tau_1) \\
    &= \frac{1-e^{i \omega_m t}+ i \omega_m t}{\omega_m^2}.
\end{align}
The terms $\beta_{B\bar{B}}(t)$ and $\beta_{B\bar{B}}^*(t)$ contain the secular terms in $t$, leading to an effective Hamiltonian description. Keeping only the secular terms from the magnetic field terms gives
\begin{align}
    \tilde{U}(t) = \mathds{1} - i \sum_{k,j=\{1,N\}}\sum_{m} \mathcal{B}_{km} \mathcal{B}_{jm} \bm{\sigma}_k^\prime \cdot\bm{\delta B}_k   \bm{\sigma}_j^\prime \cdot\bm{\delta B}_j \frac{1}{\omega_m}\left( a_m a_m^\dagger  - a_m^\dagger a_m \right)t + \dots,
\end{align}
using the canonical bosonic commutation relation $[a_m, a_l^\dagger ]=\delta_{ml}$, gives the effective Hamiltonian 
\begin{align}
    H_{\textrm{eff}} = \sum_{k,j=\{1,N\}}\sum_{m} \frac{\mathcal{B}_{km} \mathcal{B}_{jm}}{\omega_m} \bm{\sigma}_k^\prime \cdot\bm{\delta B}_k   \bm{\sigma}_j^\prime \cdot\bm{\delta B}_j.
\end{align}
In a simplified case,
$\bm{\delta B}_j = \frac{\partial \bm{B}_j}{\partial x}\big\vert_{x_j=0} = (0,0,\delta B_j)^\intercal$, so
\begin{align}
    H_{\textrm{eff}} &= \sum_{k,j=\{1,N\}}\sum_{m} \frac{\mathcal{B}_{km} \mathcal{B}_{jm} \delta B_k \delta B_j }{\omega_m} \sigma^z_k\sigma^z_j \\
     &= \sum_{k,j=\{1,N\}}\sum_{m} \frac{ g^2 \mu_B^2 b_{k m} b_{j m}\delta B_k \delta B_j }{8 m_\textrm{e} \omega_m^2} \sigma^z_k\sigma^z_j 
\end{align}
where, if we neglect the global phase, we find an effective Hamiltonian
\begin{align}
    H_{\textrm{Ising}} &= \sum_{m} \frac{ g^2 \mu_B^2 b_{1 m} b_{N m}\delta B_1 \delta B_N }{4 m_\textrm{e} \omega_m^2} \sigma^z_1\sigma^z_N.
\end{align}


\begin{thebibliography}{44}%
	\makeatletter
	\providecommand \@ifxundefined [1]{%
		\@ifx{#1\undefined}
	}%
	\providecommand \@ifnum [1]{%
		\ifnum #1\expandafter \@firstoftwo
		\else \expandafter \@secondoftwo
		\fi
	}%
	\providecommand \@ifx [1]{%
		\ifx #1\expandafter \@firstoftwo
		\else \expandafter \@secondoftwo
		\fi
	}%
	\providecommand \natexlab [1]{#1}%
	\providecommand \enquote  [1]{``#1''}%
	\providecommand \bibnamefont  [1]{#1}%
	\providecommand \bibfnamefont [1]{#1}%
	\providecommand \citenamefont [1]{#1}%
	\providecommand \href@noop [0]{\@secondoftwo}%
	\providecommand \href [0]{\begingroup \@sanitize@url \@href}%
	\providecommand \@href[1]{\@@startlink{#1}\@@href}%
	\providecommand \@@href[1]{\endgroup#1\@@endlink}%
	\providecommand \@sanitize@url [0]{\catcode `\\12\catcode `\$12\catcode
		`\&12\catcode `\#12\catcode `\^12\catcode `\_12\catcode `\%12\relax}%
	\providecommand \@@startlink[1]{}%
	\providecommand \@@endlink[0]{}%
	\providecommand \url  [0]{\begingroup\@sanitize@url \@url }%
	\providecommand \@url [1]{\endgroup\@href {#1}{\urlprefix }}%
	\providecommand \urlprefix  [0]{URL }%
	\providecommand \Eprint [0]{\href }%
	\providecommand \doibase [0]{https://doi.org/}%
	\providecommand \selectlanguage [0]{\@gobble}%
	\providecommand \bibinfo  [0]{\@secondoftwo}%
	\providecommand \bibfield  [0]{\@secondoftwo}%
	\providecommand \translation [1]{[#1]}%
	\providecommand \BibitemOpen [0]{}%
	\providecommand \bibitemStop [0]{}%
	\providecommand \bibitemNoStop [0]{.\EOS\space}%
	\providecommand \EOS [0]{\spacefactor3000\relax}%
	\providecommand \BibitemShut  [1]{\csname bibitem#1\endcsname}%
	\let\auto@bib@innerbib\@empty
	\bibitem [{\citenamefont {Montanaro}(2016)}]{montanaro_quantum_2016}%
	\BibitemOpen
	\bibfield  {author} {\bibinfo {author} {\bibfnamefont {A.}~\bibnamefont
			{Montanaro}},\ }\bibfield  {title} {{\selectlanguage {english}\bibinfo {title}
			{Quantum algorithms: an overview}},\ }\href
	{https://doi.org/10.1038/npjqi.2015.23} {\bibfield  {journal} {\bibinfo
			{journal} {npj Quantum Information}\ }\textbf {\bibinfo {volume} {2}},\
		\bibinfo {pages} {1} (\bibinfo {year} {2016})},\ \bibinfo {note} {number: 1
		Publisher: Nature Publishing Group}\BibitemShut {NoStop}%
	\bibitem [{\citenamefont {García~de Arquer}\ \emph {et~al.}(2021)\citenamefont
		{García~de Arquer}, \citenamefont {Talapin}, \citenamefont {Klimov},
		\citenamefont {Arakawa}, \citenamefont {Bayer},\ and\ \citenamefont
		{Sargent}}]{garciadearquerSemiconductorQuantumDots2021}%
	\BibitemOpen
	\bibfield  {author} {\bibinfo {author} {\bibfnamefont {F.~P.}\ \bibnamefont
			{García~de Arquer}}, \bibinfo {author} {\bibfnamefont {D.~V.}\ \bibnamefont
			{Talapin}}, \bibinfo {author} {\bibfnamefont {V.~I.}\ \bibnamefont {Klimov}},
		\bibinfo {author} {\bibfnamefont {Y.}~\bibnamefont {Arakawa}}, \bibinfo
		{author} {\bibfnamefont {M.}~\bibnamefont {Bayer}},\ and\ \bibinfo {author}
		{\bibfnamefont {E.~H.}\ \bibnamefont {Sargent}},\ }\bibfield  {title}
	{\bibinfo {title} {Semiconductor quantum dots: {Technological} progress and
			future challenges},\ }\href {https://doi.org/10.1126/science.aaz8541}
	{\bibfield  {journal} {\bibinfo  {journal} {Science}\ }\textbf {\bibinfo
			{volume} {373}},\ \bibinfo {pages} {eaaz8541} (\bibinfo {year}
		{2021})}\BibitemShut {NoStop}%
	\bibitem [{\citenamefont {Hendrickx}\ \emph {et~al.}(2021)\citenamefont
		{Hendrickx}, \citenamefont {Lawrie}, \citenamefont {Russ}, \citenamefont {van
			Riggelen}, \citenamefont {de~Snoo}, \citenamefont {Schouten}, \citenamefont
		{Sammak}, \citenamefont {Scappucci},\ and\ \citenamefont
		{Veldhorst}}]{hendrickx_four-qubit_2021}%
	\BibitemOpen
	\bibfield  {author} {\bibinfo {author} {\bibfnamefont {N.~W.}\ \bibnamefont
			{Hendrickx}}, \bibinfo {author} {\bibfnamefont {W.~I.~L.}\ \bibnamefont
			{Lawrie}}, \bibinfo {author} {\bibfnamefont {M.}~\bibnamefont {Russ}},
		\bibinfo {author} {\bibfnamefont {F.}~\bibnamefont {van Riggelen}}, \bibinfo
		{author} {\bibfnamefont {S.~L.}\ \bibnamefont {de~Snoo}}, \bibinfo {author}
		{\bibfnamefont {R.~N.}\ \bibnamefont {Schouten}}, \bibinfo {author}
		{\bibfnamefont {A.}~\bibnamefont {Sammak}}, \bibinfo {author} {\bibfnamefont
			{G.}~\bibnamefont {Scappucci}},\ and\ \bibinfo {author} {\bibfnamefont
			{M.}~\bibnamefont {Veldhorst}},\ }\bibfield  {title} {{\selectlanguage
			{english}\bibinfo {title} {A four-qubit germanium quantum processor}},\ }\href
	{https://doi.org/10.1038/s41586-021-03332-6} {\bibfield  {journal} {\bibinfo
			{journal} {Nature}\ }\textbf {\bibinfo {volume} {591}},\ \bibinfo {pages}
		{580} (\bibinfo {year} {2021})},\ \bibinfo {note} {number: 7851 Publisher:
		Nature Publishing Group}\BibitemShut {NoStop}%
	\bibitem [{\citenamefont {Philips}\ \emph {et~al.}(2022)\citenamefont
		{Philips}, \citenamefont {Mądzik}, \citenamefont {Amitonov}, \citenamefont
		{de~Snoo}, \citenamefont {Russ}, \citenamefont {Kalhor}, \citenamefont
		{Volk}, \citenamefont {Lawrie}, \citenamefont {Brousse}, \citenamefont
		{Tryputen}, \citenamefont {Wuetz}, \citenamefont {Sammak}, \citenamefont
		{Veldhorst}, \citenamefont {Scappucci},\ and\ \citenamefont
		{Vandersypen}}]{philips_universal_2022}%
	\BibitemOpen
	\bibfield  {author} {\bibinfo {author} {\bibfnamefont {S.~G.~J.}\
			\bibnamefont {Philips}}, \bibinfo {author} {\bibfnamefont {M.~T.}\
			\bibnamefont {M\c{a}dzik}}, \bibinfo {author} {\bibfnamefont {S.~V.}\
			\bibnamefont {Amitonov}}, \bibinfo {author} {\bibfnamefont {S.~L.}\
			\bibnamefont {de~Snoo}}, \bibinfo {author} {\bibfnamefont {M.}~\bibnamefont
			{Russ}}, \bibinfo {author} {\bibfnamefont {N.}~\bibnamefont {Kalhor}},
		\bibinfo {author} {\bibfnamefont {C.}~\bibnamefont {Volk}}, \bibinfo {author}
		{\bibfnamefont {W.~I.~L.}\ \bibnamefont {Lawrie}}, \bibinfo {author}
		{\bibfnamefont {D.}~\bibnamefont {Brousse}}, \bibinfo {author} {\bibfnamefont
			{L.}~\bibnamefont {Tryputen}}, \bibinfo {author} {\bibfnamefont {B.~P.}\
			\bibnamefont {Wuetz}}, \bibinfo {author} {\bibfnamefont {A.}~\bibnamefont
			{Sammak}}, \bibinfo {author} {\bibfnamefont {M.}~\bibnamefont {Veldhorst}},
		\bibinfo {author} {\bibfnamefont {G.}~\bibnamefont {Scappucci}},\ and\
		\bibinfo {author} {\bibfnamefont {L.~M.~K.}\ \bibnamefont {Vandersypen}},\
	}\bibfield  {title} {{\selectlanguage {english}\bibinfo {title} {Universal control
				of a six-qubit quantum processor in silicon}},\ }\href
	{https://doi.org/10.1038/s41586-022-05117-x} {\bibfield  {journal} {\bibinfo
			{journal} {Nature}\ }\textbf {\bibinfo {volume} {609}},\ \bibinfo {pages}
		{919} (\bibinfo {year} {2022})},\ \bibinfo {note} {number: 7929 Publisher:
		Nature Publishing Group}\BibitemShut {NoStop}%
	\bibitem [{\citenamefont {Cai}\ \emph {et~al.}(2019)\citenamefont {Cai},
		\citenamefont {Fogarty}, \citenamefont {Schaal}, \citenamefont {Patomäki},
		\citenamefont {Benjamin},\ and\ \citenamefont {Morton}}]{cai_silicon_2019}%
	\BibitemOpen
	\bibfield  {author} {\bibinfo {author} {\bibfnamefont {Z.}~\bibnamefont
			{Cai}}, \bibinfo {author} {\bibfnamefont {M.~A.}\ \bibnamefont {Fogarty}},
		\bibinfo {author} {\bibfnamefont {S.}~\bibnamefont {Schaal}}, \bibinfo
		{author} {\bibfnamefont {S.}~\bibnamefont {Patomäki}}, \bibinfo {author}
		{\bibfnamefont {S.~C.}\ \bibnamefont {Benjamin}},\ and\ \bibinfo {author}
		{\bibfnamefont {J.~J.~L.}\ \bibnamefont {Morton}},\ }\bibfield  {title}
	{{\selectlanguage {english}\bibinfo {title} {A {Silicon} {Surface} {Code}
				{Architecture} {Resilient} {Against} {Leakage} {Errors}}},\ }\href
	{https://doi.org/10.22331/q-2019-12-09-212} {\bibfield  {journal} {\bibinfo
			{journal} {Quantum}\ }\textbf {\bibinfo {volume} {3}},\ \bibinfo {pages}
		{212} (\bibinfo {year} {2019})},\ \bibinfo {note} {publisher: Verein zur
		Förderung des Open Access Publizierens in den
		Quantenwissenschaften}\BibitemShut {NoStop}%
	\bibitem [{\citenamefont {Jnane}\ \emph {et~al.}(2022)\citenamefont {Jnane},
		\citenamefont {Undseth}, \citenamefont {Cai}, \citenamefont {Benjamin},\ and\
		\citenamefont {Koczor}}]{jnane_multicore_2022}%
	\BibitemOpen
	\bibfield  {author} {\bibinfo {author} {\bibfnamefont {H.}~\bibnamefont
			{Jnane}}, \bibinfo {author} {\bibfnamefont {B.}~\bibnamefont {Undseth}},
		\bibinfo {author} {\bibfnamefont {Z.}~\bibnamefont {Cai}}, \bibinfo {author}
		{\bibfnamefont {S.~C.}\ \bibnamefont {Benjamin}},\ and\ \bibinfo {author}
		{\bibfnamefont {B.}~\bibnamefont {Koczor}},\ }\bibfield  {title} {\bibinfo
		{title} {Multicore {Quantum} {Computing}},\ }\href
	{https://doi.org/10.1103/PhysRevApplied.18.044064} {\bibfield  {journal}
		{\bibinfo  {journal} {Physical Review Applied}\ }\textbf {\bibinfo {volume}
			{18}},\ \bibinfo {pages} {044064} (\bibinfo {year} {2022})},\ \bibinfo {note}
	{publisher: American Physical Society}\BibitemShut {NoStop}%
	\bibitem [{\citenamefont {Boter}\ \emph {et~al.}(2022)\citenamefont {Boter},
		\citenamefont {Dehollain}, \citenamefont {van Dijk}, \citenamefont {Xu},
		\citenamefont {Hensgens}, \citenamefont {Versluis}, \citenamefont {Naus},
		\citenamefont {Clarke}, \citenamefont {Veldhorst}, \citenamefont
		{Sebastiano},\ and\ \citenamefont {Vandersypen}}]{boter_spiderweb_2022}%
	\BibitemOpen
	\bibfield  {author} {\bibinfo {author} {\bibfnamefont {J.~M.}\ \bibnamefont
			{Boter}}, \bibinfo {author} {\bibfnamefont {J.~P.}\ \bibnamefont
			{Dehollain}}, \bibinfo {author} {\bibfnamefont {J.~P.}\ \bibnamefont {van
				Dijk}}, \bibinfo {author} {\bibfnamefont {Y.}~\bibnamefont {Xu}}, \bibinfo
		{author} {\bibfnamefont {T.}~\bibnamefont {Hensgens}}, \bibinfo {author}
		{\bibfnamefont {R.}~\bibnamefont {Versluis}}, \bibinfo {author}
		{\bibfnamefont {H.~W.}\ \bibnamefont {Naus}}, \bibinfo {author}
		{\bibfnamefont {J.~S.}\ \bibnamefont {Clarke}}, \bibinfo {author}
		{\bibfnamefont {M.}~\bibnamefont {Veldhorst}}, \bibinfo {author}
		{\bibfnamefont {F.}~\bibnamefont {Sebastiano}},\ and\ \bibinfo {author}
		{\bibfnamefont {L.~M.}\ \bibnamefont {Vandersypen}},\ }\bibfield  {title}
	{\bibinfo {title} {Spiderweb {Array}: {A} {Sparse} {Spin}-{Qubit} {Array}},\
	}\href {https://doi.org/10.1103/PhysRevApplied.18.024053} {\bibfield
		{journal} {\bibinfo  {journal} {Physical Review Applied}\ }\textbf {\bibinfo
			{volume} {18}},\ \bibinfo {pages} {024053} (\bibinfo {year} {2022})},\
	\bibinfo {note} {publisher: American Physical Society}\BibitemShut {NoStop}%
	\bibitem [{\citenamefont {Patomäki}\ \emph {et~al.}(2023)\citenamefont
		{Patomäki}, \citenamefont {Gonzalez-Zalba}, \citenamefont {Fogarty},
		\citenamefont {Cai}, \citenamefont {Benjamin},\ and\ \citenamefont
		{Morton}}]{patomaki_pipeline_2023}%
	\BibitemOpen
	\bibfield  {author} {\bibinfo {author} {\bibfnamefont {S.~M.}\ \bibnamefont
			{Patomäki}}, \bibinfo {author} {\bibfnamefont {M.~F.}\ \bibnamefont
			{Gonzalez-Zalba}}, \bibinfo {author} {\bibfnamefont {M.~A.}\ \bibnamefont
			{Fogarty}}, \bibinfo {author} {\bibfnamefont {Z.}~\bibnamefont {Cai}},
		\bibinfo {author} {\bibfnamefont {S.~C.}\ \bibnamefont {Benjamin}},\ and\
		\bibinfo {author} {\bibfnamefont {J.~J.~L.}\ \bibnamefont {Morton}},\ }\href
	{https://doi.org/10.48550/arXiv.2306.07673} {\bibinfo {title} {Pipeline
			quantum processor architecture for silicon spin qubits}} (\bibinfo {year}
	{2023}),\ \bibinfo {note} {arXiv:2306.07673 [quant-ph]}\BibitemShut {NoStop}%
	\bibitem [{\citenamefont {Baart}\ \emph {et~al.}(2016)\citenamefont {Baart},
		\citenamefont {Shafiei}, \citenamefont {Fujita}, \citenamefont {Reichl},
		\citenamefont {Wegscheider},\ and\ \citenamefont
		{Vandersypen}}]{baart_single-spin_2016}%
	\BibitemOpen
	\bibfield  {author} {\bibinfo {author} {\bibfnamefont {T.~A.}\ \bibnamefont
			{Baart}}, \bibinfo {author} {\bibfnamefont {M.}~\bibnamefont {Shafiei}},
		\bibinfo {author} {\bibfnamefont {T.}~\bibnamefont {Fujita}}, \bibinfo
		{author} {\bibfnamefont {C.}~\bibnamefont {Reichl}}, \bibinfo {author}
		{\bibfnamefont {W.}~\bibnamefont {Wegscheider}},\ and\ \bibinfo {author}
		{\bibfnamefont {L.~M.~K.}\ \bibnamefont {Vandersypen}},\ }\bibfield  {title}
	{{\selectlanguage {english}\bibinfo {title} {Single-spin {CCD}}},\ }\href
	{https://doi.org/10.1038/nnano.2015.291} {\bibfield  {journal} {\bibinfo
			{journal} {Nature Nanotechnology}\ }\textbf {\bibinfo {volume} {11}},\
		\bibinfo {pages} {330} (\bibinfo {year} {2016})},\ \bibinfo {note}
	{bandiera\_abtest: a Cg\_type: Nature Research Journals Number: 4
		Primary\_atype: Research Publisher: Nature Publishing Group Subject\_term:
		Electronic and spintronic devices;Nanoscale devices;Qubits;Spintronics
		Subject\_term\_id:
		electronic-and-spintronic-devices;nanoscale-devices;qubits;spintronics}\BibitemShut
	{NoStop}%
	\bibitem [{\citenamefont {Fujita}\ \emph {et~al.}(2017)\citenamefont {Fujita},
		\citenamefont {Baart}, \citenamefont {Reichl}, \citenamefont {Wegscheider},\
		and\ \citenamefont {Vandersypen}}]{fujita_coherent_2017}%
	\BibitemOpen
	\bibfield  {author} {\bibinfo {author} {\bibfnamefont {T.}~\bibnamefont
			{Fujita}}, \bibinfo {author} {\bibfnamefont {T.~A.}\ \bibnamefont {Baart}},
		\bibinfo {author} {\bibfnamefont {C.}~\bibnamefont {Reichl}}, \bibinfo
		{author} {\bibfnamefont {W.}~\bibnamefont {Wegscheider}},\ and\ \bibinfo
		{author} {\bibfnamefont {L.~M.~K.}\ \bibnamefont {Vandersypen}},\ }\bibfield
	{title} {{\selectlanguage {english}\bibinfo {title} {Coherent shuttle of
				electron-spin states}},\ }\href {https://doi.org/10.1038/s41534-017-0024-4}
	{\bibfield  {journal} {\bibinfo  {journal} {npj Quantum Information}\
		}\textbf {\bibinfo {volume} {3}},\ \bibinfo {pages} {1} (\bibinfo {year}
		{2017})},\ \bibinfo {note} {number: 1 Publisher: Nature Publishing
		Group}\BibitemShut {NoStop}%
	\bibitem [{\citenamefont {Mills}\ \emph {et~al.}(2019)\citenamefont {Mills},
		\citenamefont {Zajac}, \citenamefont {Gullans}, \citenamefont {Schupp},
		\citenamefont {Hazard},\ and\ \citenamefont {Petta}}]{mills_shuttling_2019}%
	\BibitemOpen
	\bibfield  {author} {\bibinfo {author} {\bibfnamefont {A.~R.}\ \bibnamefont
			{Mills}}, \bibinfo {author} {\bibfnamefont {D.~M.}\ \bibnamefont {Zajac}},
		\bibinfo {author} {\bibfnamefont {M.~J.}\ \bibnamefont {Gullans}}, \bibinfo
		{author} {\bibfnamefont {F.~J.}\ \bibnamefont {Schupp}}, \bibinfo {author}
		{\bibfnamefont {T.~M.}\ \bibnamefont {Hazard}},\ and\ \bibinfo {author}
		{\bibfnamefont {J.~R.}\ \bibnamefont {Petta}},\ }\bibfield  {title}
	{{\selectlanguage {english}\bibinfo {title} {Shuttling a single charge across a
				one-dimensional array of silicon quantum dots}},\ }\href
	{https://doi.org/10.1038/s41467-019-08970-z} {\bibfield  {journal} {\bibinfo
			{journal} {Nature Communications}\ }\textbf {\bibinfo {volume} {10}},\
		\bibinfo {pages} {1063} (\bibinfo {year} {2019})},\ \bibinfo {note}
	{bandiera\_abtest: a Cc\_license\_type: cc\_by Cg\_type: Nature Research
		Journals Number: 1 Primary\_atype: Research Publisher: Nature Publishing
		Group Subject\_term: Electronic devices;Quantum dots;Quantum
		information;Qubits Subject\_term\_id:
		electronic-devices;quantum-dots;quantum-information;qubits}\BibitemShut
	{NoStop}%
	\bibitem [{\citenamefont {Buonacorsi}\ \emph {et~al.}(2020)\citenamefont
		{Buonacorsi}, \citenamefont {Shaw},\ and\ \citenamefont
		{Baugh}}]{buonacorsi_simulated_2020}%
	\BibitemOpen
	\bibfield  {author} {\bibinfo {author} {\bibfnamefont {B.}~\bibnamefont
			{Buonacorsi}}, \bibinfo {author} {\bibfnamefont {B.}~\bibnamefont {Shaw}},\
		and\ \bibinfo {author} {\bibfnamefont {J.}~\bibnamefont {Baugh}},\ }\bibfield
	{title} {\bibinfo {title} {Simulated coherent electron shuttling in silicon
			quantum dots},\ }\href {https://doi.org/10.1103/PhysRevB.102.125406}
	{\bibfield  {journal} {\bibinfo  {journal} {Physical Review B}\ }\textbf
		{\bibinfo {volume} {102}},\ \bibinfo {pages} {125406} (\bibinfo {year}
		{2020})},\ \bibinfo {note} {publisher: American Physical Society}\BibitemShut
	{NoStop}%
	\bibitem [{\citenamefont {Ginzel}\ \emph {et~al.}(2020)\citenamefont {Ginzel},
		\citenamefont {Mills}, \citenamefont {Petta},\ and\ \citenamefont
		{Burkard}}]{ginzel_spin_2020}%
	\BibitemOpen
	\bibfield  {author} {\bibinfo {author} {\bibfnamefont {F.}~\bibnamefont
			{Ginzel}}, \bibinfo {author} {\bibfnamefont {A.~R.}\ \bibnamefont {Mills}},
		\bibinfo {author} {\bibfnamefont {J.~R.}\ \bibnamefont {Petta}},\ and\
		\bibinfo {author} {\bibfnamefont {G.}~\bibnamefont {Burkard}},\ }\bibfield
	{title} {\bibinfo {title} {Spin shuttling in a silicon double quantum dot},\
	}\href {https://doi.org/10.1103/PhysRevB.102.195418} {\bibfield  {journal}
		{\bibinfo  {journal} {Physical Review B}\ }\textbf {\bibinfo {volume}
			{102}},\ \bibinfo {pages} {195418} (\bibinfo {year} {2020})},\ \bibinfo
	{note} {publisher: American Physical Society}\BibitemShut {NoStop}%
	\bibitem [{\citenamefont {Seidler}\ \emph {et~al.}(2021)\citenamefont
		{Seidler}, \citenamefont {Struck}, \citenamefont {Xue}, \citenamefont
		{Focke}, \citenamefont {Trellenkamp}, \citenamefont {Bluhm},\ and\
		\citenamefont {Schreiber}}]{seidler_conveyor-mode_2021}%
	\BibitemOpen
	\bibfield  {author} {\bibinfo {author} {\bibfnamefont {I.}~\bibnamefont
			{Seidler}}, \bibinfo {author} {\bibfnamefont {T.}~\bibnamefont {Struck}},
		\bibinfo {author} {\bibfnamefont {R.}~\bibnamefont {Xue}}, \bibinfo {author}
		{\bibfnamefont {N.}~\bibnamefont {Focke}}, \bibinfo {author} {\bibfnamefont
			{S.}~\bibnamefont {Trellenkamp}}, \bibinfo {author} {\bibfnamefont
			{H.}~\bibnamefont {Bluhm}},\ and\ \bibinfo {author} {\bibfnamefont {L.~R.}\
			\bibnamefont {Schreiber}},\ }\bibfield  {title} {\bibinfo {title}
		{Conveyor-mode single-electron shuttling in {Si}/{SiGe} for a scalable
			quantum computing architecture},\ }\href {http://arxiv.org/abs/2108.00879}
	{\bibfield  {journal} {\bibinfo  {journal} {arXiv:2108.00879 [cond-mat,
				physics:quant-ph]}\ } (\bibinfo {year} {2021})},\ \bibinfo {note} {arXiv:
		2108.00879}\BibitemShut {NoStop}%
	\bibitem [{\citenamefont {Borsoi}\ \emph {et~al.}(2023)\citenamefont {Borsoi},
		\citenamefont {Hendrickx}, \citenamefont {John}, \citenamefont {Meyer},
		\citenamefont {Motz}, \citenamefont {van Riggelen}, \citenamefont {Sammak},
		\citenamefont {de~Snoo}, \citenamefont {Scappucci},\ and\ \citenamefont
		{Veldhorst}}]{borsoi_shared_2023}%
	\BibitemOpen
	\bibfield  {author} {\bibinfo {author} {\bibfnamefont {F.}~\bibnamefont
			{Borsoi}}, \bibinfo {author} {\bibfnamefont {N.~W.}\ \bibnamefont
			{Hendrickx}}, \bibinfo {author} {\bibfnamefont {V.}~\bibnamefont {John}},
		\bibinfo {author} {\bibfnamefont {M.}~\bibnamefont {Meyer}}, \bibinfo
		{author} {\bibfnamefont {S.}~\bibnamefont {Motz}}, \bibinfo {author}
		{\bibfnamefont {F.}~\bibnamefont {van Riggelen}}, \bibinfo {author}
		{\bibfnamefont {A.}~\bibnamefont {Sammak}}, \bibinfo {author} {\bibfnamefont
			{S.~L.}\ \bibnamefont {de~Snoo}}, \bibinfo {author} {\bibfnamefont
			{G.}~\bibnamefont {Scappucci}},\ and\ \bibinfo {author} {\bibfnamefont
			{M.}~\bibnamefont {Veldhorst}},\ }\bibfield  {title} {{\selectlanguage
			{english}\bibinfo {title} {Shared control of a 16 semiconductor quantum dot
				crossbar array}},\ }\href {https://doi.org/10.1038/s41565-023-01491-3}
	{\bibfield  {journal} {\bibinfo  {journal} {Nature Nanotechnology}\ ,\
			\bibinfo {pages} {1}} (\bibinfo {year} {2023})},\ \bibinfo {note} {publisher:
		Nature Publishing Group}\BibitemShut {NoStop}%
	\bibitem [{\citenamefont {Bose}(2003)}]{Bose2003}%
	\BibitemOpen
	\bibfield  {author} {\bibinfo {author} {\bibfnamefont {S.}~\bibnamefont
			{Bose}},\ }\bibfield  {title} {\bibinfo {title} {Quantum communication
			through an unmodulated spin chain},\ }\href
	{https://doi.org/10.1103/PhysRevLett.91.207901} {\bibfield  {journal}
		{\bibinfo  {journal} {Physical Review Letters}\ }\textbf {\bibinfo {volume}
			{91}},\ \bibinfo {pages} {207901} (\bibinfo {year} {2003})}\BibitemShut
	{NoStop}%
	\bibitem [{\citenamefont {Antonio}\ \emph {et~al.}(2015)\citenamefont
		{Antonio}, \citenamefont {Bayat}, \citenamefont {Kumar}, \citenamefont
		{Pepper},\ and\ \citenamefont
		{Bose}}]{antonioSelfAssembledWignerCrystals2015}%
	\BibitemOpen
	\bibfield  {author} {\bibinfo {author} {\bibfnamefont {B.}~\bibnamefont
			{Antonio}}, \bibinfo {author} {\bibfnamefont {A.}~\bibnamefont {Bayat}},
		\bibinfo {author} {\bibfnamefont {S.}~\bibnamefont {Kumar}}, \bibinfo
		{author} {\bibfnamefont {M.}~\bibnamefont {Pepper}},\ and\ \bibinfo {author}
		{\bibfnamefont {S.}~\bibnamefont {Bose}},\ }\bibfield  {title} {\bibinfo
		{title} {Self-{Assembled} {Wigner} {Crystals} as {Mediators} of {Spin}
			{Currents} and {Quantum} {Information}},\ }\href
	{https://doi.org/10.1103/PhysRevLett.115.216804} {\bibfield  {journal}
		{\bibinfo  {journal} {Physical Review Letters}\ }\textbf {\bibinfo {volume}
			{115}},\ \bibinfo {pages} {216804} (\bibinfo {year} {2015})}\BibitemShut
	{NoStop}%
	\bibitem [{\citenamefont {Lewis}\ \emph
		{et~al.}(2023{\natexlab{a}})\citenamefont {Lewis}, \citenamefont {Moutinho},
		\citenamefont {Costa}, \citenamefont {Omar},\ and\ \citenamefont
		{Bose}}]{lewis_low-dissipation_2023}%
	\BibitemOpen
	\bibfield  {author} {\bibinfo {author} {\bibfnamefont {D.}~\bibnamefont
			{Lewis}}, \bibinfo {author} {\bibfnamefont {J.~P.}\ \bibnamefont {Moutinho}},
		\bibinfo {author} {\bibfnamefont {A.~T.}\ \bibnamefont {Costa}}, \bibinfo
		{author} {\bibfnamefont {Y.}~\bibnamefont {Omar}},\ and\ \bibinfo {author}
		{\bibfnamefont {S.}~\bibnamefont {Bose}},\ }\bibfield  {title} {\bibinfo
		{title} {Low-dissipation data bus via coherent quantum dynamics},\ }\href
	{https://doi.org/10.1103/PhysRevB.108.075405} {\bibfield  {journal} {\bibinfo
			{journal} {Physical Review B}\ }\textbf {\bibinfo {volume} {108}},\ \bibinfo
		{pages} {075405} (\bibinfo {year} {2023}{\natexlab{a}})},\ \bibinfo {note}
	{publisher: American Physical Society}\BibitemShut {NoStop}%
	\bibitem [{\citenamefont {Orth}\ \emph {et~al.}(2008)\citenamefont {Orth},
		\citenamefont {Stanic},\ and\ \citenamefont
		{Le~Hur}}]{orthDissipativeQuantumIsing2008}%
	\BibitemOpen
	\bibfield  {author} {\bibinfo {author} {\bibfnamefont {P.~P.}\ \bibnamefont
			{Orth}}, \bibinfo {author} {\bibfnamefont {I.}~\bibnamefont {Stanic}},\ and\
		\bibinfo {author} {\bibfnamefont {K.}~\bibnamefont {Le~Hur}},\ }\bibfield
	{title} {\bibinfo {title} {Dissipative quantum {Ising} model in a cold-atom
			spin-boson mixture},\ }\href {https://doi.org/10.1103/PhysRevA.77.051601}
	{\bibfield  {journal} {\bibinfo  {journal} {Physical Review A}\ }\textbf
		{\bibinfo {volume} {77}},\ \bibinfo {pages} {051601} (\bibinfo {year}
		{2008})}\BibitemShut {NoStop}%
	\bibitem [{\citenamefont {Orth}\ \emph {et~al.}(2010)\citenamefont {Orth},
		\citenamefont {Roosen}, \citenamefont {Hofstetter},\ and\ \citenamefont
		{Le~Hur}}]{orthDynamicsSynchronizationQuantum2010}%
	\BibitemOpen
	\bibfield  {author} {\bibinfo {author} {\bibfnamefont {P.~P.}\ \bibnamefont
			{Orth}}, \bibinfo {author} {\bibfnamefont {D.}~\bibnamefont {Roosen}},
		\bibinfo {author} {\bibfnamefont {W.}~\bibnamefont {Hofstetter}},\ and\
		\bibinfo {author} {\bibfnamefont {K.}~\bibnamefont {Le~Hur}},\ }\bibfield
	{title} {\bibinfo {title} {Dynamics, synchronization, and quantum phase
			transitions of two dissipative spins},\ }\href
	{https://doi.org/10.1103/PhysRevB.82.144423} {\bibfield  {journal} {\bibinfo
			{journal} {Physical Review B}\ }\textbf {\bibinfo {volume} {82}},\ \bibinfo
		{pages} {144423} (\bibinfo {year} {2010})}\BibitemShut {NoStop}%
	\bibitem [{\citenamefont {Wall}\ \emph {et~al.}(2017)\citenamefont {Wall},
		\citenamefont {Safavi-Naini},\ and\ \citenamefont
		{Rey}}]{wallBosonmediatedQuantumSpin2017}%
	\BibitemOpen
	\bibfield  {author} {\bibinfo {author} {\bibfnamefont {M.~L.}\ \bibnamefont
			{Wall}}, \bibinfo {author} {\bibfnamefont {A.}~\bibnamefont {Safavi-Naini}},\
		and\ \bibinfo {author} {\bibfnamefont {A.~M.}\ \bibnamefont {Rey}},\
	}\bibfield  {title} {\bibinfo {title} {Boson-mediated quantum spin simulators
			in transverse fields: \${XY}\$ model and spin-boson entanglement},\ }\href
	{https://doi.org/10.1103/PhysRevA.95.013602} {\bibfield  {journal} {\bibinfo
			{journal} {Physical Review A}\ }\textbf {\bibinfo {volume} {95}},\ \bibinfo
		{pages} {013602} (\bibinfo {year} {2017})}\BibitemShut {NoStop}%
	\bibitem [{\citenamefont {Lewis}\ \emph
		{et~al.}(2023{\natexlab{b}})\citenamefont {Lewis}, \citenamefont {Banchi},
		\citenamefont {Teoh}, \citenamefont {Islam},\ and\ \citenamefont
		{Bose}}]{lewisIonTrapLongrange2023}%
	\BibitemOpen
	\bibfield  {author} {\bibinfo {author} {\bibfnamefont {D.}~\bibnamefont
			{Lewis}}, \bibinfo {author} {\bibfnamefont {L.}~\bibnamefont {Banchi}},
		\bibinfo {author} {\bibfnamefont {Y.~H.}\ \bibnamefont {Teoh}}, \bibinfo
		{author} {\bibfnamefont {R.}~\bibnamefont {Islam}},\ and\ \bibinfo {author}
		{\bibfnamefont {S.}~\bibnamefont {Bose}},\ }\bibfield  {title}
	{{\selectlanguage {english}\bibinfo {title} {Ion trap long-range {XY} model for
				quantum state transfer and optimal spatial search}},\ }\href
	{https://doi.org/10.1088/2058-9565/acd953} {\bibfield  {journal} {\bibinfo
			{journal} {Quantum Science and Technology}\ }\textbf {\bibinfo {volume}
			{8}},\ \bibinfo {pages} {035025} (\bibinfo {year}
		{2023}{\natexlab{b}})}\BibitemShut {NoStop}%
	\bibitem [{\citenamefont {Bernhardt}\ \emph {et~al.}(2024)\citenamefont
		{Bernhardt}, \citenamefont {Cheung},\ and\ \citenamefont
		{Le~Hur}}]{bernhardtMajoranaFermionsQuantum2024}%
	\BibitemOpen
	\bibfield  {author} {\bibinfo {author} {\bibfnamefont {E.}~\bibnamefont
			{Bernhardt}}, \bibinfo {author} {\bibfnamefont {B.~C.~H.}\ \bibnamefont
			{Cheung}},\ and\ \bibinfo {author} {\bibfnamefont {K.}~\bibnamefont
			{Le~Hur}},\ }\bibfield  {title} {\bibinfo {title} {Majorana fermions and
			quantum information with fractional topology and disorder},\ }\href
	{https://doi.org/10.1103/PhysRevResearch.6.023221} {\bibfield  {journal}
		{\bibinfo  {journal} {Physical Review Research}\ }\textbf {\bibinfo {volume}
			{6}},\ \bibinfo {pages} {023221} (\bibinfo {year} {2024})}\BibitemShut
	{NoStop}%
	\bibitem [{\citenamefont {Huang}\ \emph {et~al.}(2025)\citenamefont {Huang},
		\citenamefont {Hausten}, \citenamefont {Yu}, \citenamefont {Taniguchi},
		\citenamefont {Yadav}, \citenamefont {Sacksteder}, \citenamefont {Noguchi},
		\citenamefont {Schneider},\ and\ \citenamefont
		{Haeffner}}]{huangNumericalInvestigationsElectron2025}%
	\BibitemOpen
	\bibfield  {author} {\bibinfo {author} {\bibfnamefont {A.}~\bibnamefont
			{Huang}}, \bibinfo {author} {\bibfnamefont {E.}~\bibnamefont {Hausten}},
		\bibinfo {author} {\bibfnamefont {Q.}~\bibnamefont {Yu}}, \bibinfo {author}
		{\bibfnamefont {K.}~\bibnamefont {Taniguchi}}, \bibinfo {author}
		{\bibfnamefont {N.}~\bibnamefont {Yadav}}, \bibinfo {author} {\bibfnamefont
			{I.}~\bibnamefont {Sacksteder}}, \bibinfo {author} {\bibfnamefont
			{A.}~\bibnamefont {Noguchi}}, \bibinfo {author} {\bibfnamefont
			{R.}~\bibnamefont {Schneider}},\ and\ \bibinfo {author} {\bibfnamefont
			{H.}~\bibnamefont {Haeffner}},\ }\href
	{https://doi.org/10.48550/arXiv.2503.12379} {\bibinfo {title} {Numerical
			{Investigations} of {Electron} {Dynamics} in a {Linear} {Paul} {Trap}}}
	(\bibinfo {year} {2025}),\ \bibinfo {note} {arXiv:2503.12379
		[quant-ph]}\BibitemShut {NoStop}%
	\bibitem [{\citenamefont {Wensauer}\ \emph {et~al.}(2000)\citenamefont
		{Wensauer}, \citenamefont {Steffens}, \citenamefont {Suhrke},\ and\
		\citenamefont {Rössler}}]{wensauer_laterally_2000}%
	\BibitemOpen
	\bibfield  {author} {\bibinfo {author} {\bibfnamefont {A.}~\bibnamefont
			{Wensauer}}, \bibinfo {author} {\bibfnamefont {O.}~\bibnamefont {Steffens}},
		\bibinfo {author} {\bibfnamefont {M.}~\bibnamefont {Suhrke}},\ and\ \bibinfo
		{author} {\bibfnamefont {U.}~\bibnamefont {Rössler}},\ }\bibfield  {title}
	{{\selectlanguage {english}\bibinfo {title} {Laterally coupled few-electron
				quantum dots}},\ }\href {https://doi.org/10.1103/PhysRevB.62.2605} {\bibfield
		{journal} {\bibinfo  {journal} {Physical Review B}\ }\textbf {\bibinfo
			{volume} {62}},\ \bibinfo {pages} {2605} (\bibinfo {year}
		{2000})}\BibitemShut {NoStop}%
	\bibitem [{\citenamefont {Helle}\ \emph {et~al.}(2005)\citenamefont {Helle},
		\citenamefont {Harju},\ and\ \citenamefont
		{Nieminen}}]{helle_two-electron_2005}%
	\BibitemOpen
	\bibfield  {author} {\bibinfo {author} {\bibfnamefont {M.}~\bibnamefont
			{Helle}}, \bibinfo {author} {\bibfnamefont {A.}~\bibnamefont {Harju}},\ and\
		\bibinfo {author} {\bibfnamefont {R.~M.}\ \bibnamefont {Nieminen}},\
	}\bibfield  {title} {{\selectlanguage {english}\bibinfo {title} {Two-electron
				lateral quantum-dot molecules in a magnetic field}},\ }\href
	{https://doi.org/10.1103/PhysRevB.72.205329} {\bibfield  {journal} {\bibinfo
			{journal} {Physical Review B}\ }\textbf {\bibinfo {volume} {72}},\ \bibinfo
		{pages} {205329} (\bibinfo {year} {2005})}\BibitemShut {NoStop}%
	\bibitem [{\citenamefont {Li}\ \emph {et~al.}(2010)\citenamefont {Li},
		\citenamefont {Cywi\'{n}ski}, \citenamefont {Culcer}, \citenamefont {Hu},\ and\
		\citenamefont {Das~Sarma}}]{li_exchange_2010}%
	\BibitemOpen
	\bibfield  {author} {\bibinfo {author} {\bibfnamefont {Q.}~\bibnamefont
			{Li}}, \bibinfo {author} {\bibfnamefont {\L.}~\bibnamefont {Cywiński}},
		\bibinfo {author} {\bibfnamefont {D.}~\bibnamefont {Culcer}}, \bibinfo
		{author} {\bibfnamefont {X.}~\bibnamefont {Hu}},\ and\ \bibinfo {author}
		{\bibfnamefont {S.}~\bibnamefont {Das Sarma}},\ }\bibfield  {title}
	{{\selectlanguage {english}\bibinfo {title} {Exchange coupling in silicon quantum
				dots: {Theoretical} considerations for quantum computation}},\ }\href
	{https://doi.org/10.1103/PhysRevB.81.085313} {\bibfield  {journal} {\bibinfo
			{journal} {Physical Review B}\ }\textbf {\bibinfo {volume} {81}},\ \bibinfo
		{pages} {085313} (\bibinfo {year} {2010})}\BibitemShut {NoStop}%
	\bibitem [{\citenamefont {Yang}\ \emph {et~al.}(2011)\citenamefont {Yang},
		\citenamefont {Wang},\ and\ \citenamefont {Das~Sarma}}]{yang_generic_2011}%
	\BibitemOpen
	\bibfield  {author} {\bibinfo {author} {\bibfnamefont {S.}~\bibnamefont
			{Yang}}, \bibinfo {author} {\bibfnamefont {X.}~\bibnamefont {Wang}},\ and\
		\bibinfo {author} {\bibfnamefont {S.}~\bibnamefont {Das~Sarma}},\ }\bibfield
	{title} {{\selectlanguage {english}\bibinfo {title} {Generic {Hubbard} model
				description of semiconductor quantum-dot spin qubits}},\ }\href
	{https://doi.org/10.1103/PhysRevB.83.161301} {\bibfield  {journal} {\bibinfo
			{journal} {Physical Review B}\ }\textbf {\bibinfo {volume} {83}},\ \bibinfo
		{pages} {161301} (\bibinfo {year} {2011})}\BibitemShut {NoStop}%
	\bibitem [{\citenamefont {Reimann}\ and\ \citenamefont
		{Manninen}(2002)}]{reimann_electronic_2002}%
	\BibitemOpen
	\bibfield  {author} {\bibinfo {author} {\bibfnamefont {S.~M.}\ \bibnamefont
			{Reimann}}\ and\ \bibinfo {author} {\bibfnamefont {M.}~\bibnamefont
			{Manninen}},\ }\bibfield  {title} {\bibinfo {title} {Electronic structure of
			quantum dots},\ }\href {https://doi.org/10.1103/RevModPhys.74.1283}
	{\bibfield  {journal} {\bibinfo  {journal} {Reviews of Modern Physics}\
		}\textbf {\bibinfo {volume} {74}},\ \bibinfo {pages} {1283} (\bibinfo {year}
		{2002})},\ \bibinfo {note} {publisher: American Physical Society}\BibitemShut
	{NoStop}%
	\bibitem [{\citenamefont {Fishman}\ \emph {et~al.}(2008)\citenamefont
		{Fishman}, \citenamefont {De~Chiara}, \citenamefont {Calarco},\ and\
		\citenamefont {Morigi}}]{fishman_structural_2008}%
	\BibitemOpen
	\bibfield  {author} {\bibinfo {author} {\bibfnamefont {S.}~\bibnamefont
			{Fishman}}, \bibinfo {author} {\bibfnamefont {G.}~\bibnamefont {De~Chiara}},
		\bibinfo {author} {\bibfnamefont {T.}~\bibnamefont {Calarco}},\ and\ \bibinfo
		{author} {\bibfnamefont {G.}~\bibnamefont {Morigi}},\ }\bibfield  {title}
	{\bibinfo {title} {Structural phase transitions in low-dimensional ion
			crystals},\ }\href {https://doi.org/10.1103/PhysRevB.77.064111} {\bibfield
		{journal} {\bibinfo  {journal} {Physical Review B}\ }\textbf {\bibinfo
			{volume} {77}},\ \bibinfo {pages} {064111} (\bibinfo {year} {2008})},\
	\bibinfo {note} {publisher: American Physical Society}\BibitemShut {NoStop}%
	\bibitem [{\citenamefont {Lewis}(2025)}]{our_data}%
	\BibitemOpen
	\bibfield  {author} {\bibinfo {author} {\bibfnamefont {D.}~\bibnamefont
			{Lewis}},\ }\href@noop {} {\bibinfo {title} {{Spin qubit gates via phonon
				buses in electron nanowires}}} (\bibinfo {year} {2025}),\ \bibinfo {note}
	{\url{https://github.com/dyylan/electron_nanowire}}\BibitemShut {NoStop}%
	\bibitem [{\citenamefont {Petta}\ \emph {et~al.}(2005)\citenamefont {Petta},
		\citenamefont {Johnson}, \citenamefont {Taylor}, \citenamefont {Laird},
		\citenamefont {Yacoby}, \citenamefont {Lukin}, \citenamefont {Marcus},
		\citenamefont {Hanson},\ and\ \citenamefont
		{Gossard}}]{pettaCoherentManipulationCoupled2005}%
	\BibitemOpen
	\bibfield  {author} {\bibinfo {author} {\bibfnamefont {J.~R.}\ \bibnamefont
			{Petta}}, \bibinfo {author} {\bibfnamefont {A.~C.}\ \bibnamefont {Johnson}},
		\bibinfo {author} {\bibfnamefont {J.~M.}\ \bibnamefont {Taylor}}, \bibinfo
		{author} {\bibfnamefont {E.~A.}\ \bibnamefont {Laird}}, \bibinfo {author}
		{\bibfnamefont {A.}~\bibnamefont {Yacoby}}, \bibinfo {author} {\bibfnamefont
			{M.~D.}\ \bibnamefont {Lukin}}, \bibinfo {author} {\bibfnamefont {C.~M.}\
			\bibnamefont {Marcus}}, \bibinfo {author} {\bibfnamefont {M.~P.}\
			\bibnamefont {Hanson}},\ and\ \bibinfo {author} {\bibfnamefont {A.~C.}\
			\bibnamefont {Gossard}},\ }\bibfield  {title} {\bibinfo {title} {Coherent
			{Manipulation} of {Coupled} {Electron} {Spins} in {Semiconductor} {Quantum}
			{Dots}},\ }\href {https://doi.org/10.1126/science.1116955} {\bibfield
		{journal} {\bibinfo  {journal} {Science}\ }\textbf {\bibinfo {volume}
			{309}},\ \bibinfo {pages} {2180} (\bibinfo {year} {2005})}\BibitemShut
	{NoStop}%
	\bibitem [{\citenamefont {Hanson}\ \emph {et~al.}(2003)\citenamefont {Hanson},
		\citenamefont {Witkamp}, \citenamefont {Vandersypen}, \citenamefont {van
			Beveren}, \citenamefont {Elzerman},\ and\ \citenamefont
		{Kouwenhoven}}]{hansonZeemanEnergySpin2003}%
	\BibitemOpen
	\bibfield  {author} {\bibinfo {author} {\bibfnamefont {R.}~\bibnamefont
			{Hanson}}, \bibinfo {author} {\bibfnamefont {B.}~\bibnamefont {Witkamp}},
		\bibinfo {author} {\bibfnamefont {L.~M.~K.}\ \bibnamefont {Vandersypen}},
		\bibinfo {author} {\bibfnamefont {L.~H.~W.}\ \bibnamefont {van Beveren}},
		\bibinfo {author} {\bibfnamefont {J.~M.}\ \bibnamefont {Elzerman}},\ and\
		\bibinfo {author} {\bibfnamefont {L.~P.}\ \bibnamefont {Kouwenhoven}},\
	}\bibfield  {title} {\bibinfo {title} {Zeeman {Energy} and {Spin}
			{Relaxation} in a {One}-{Electron} {Quantum} {Dot}},\ }\href
	{https://doi.org/10.1103/PhysRevLett.91.196802} {\bibfield  {journal}
		{\bibinfo  {journal} {Physical Review Letters}\ }\textbf {\bibinfo {volume}
			{91}},\ \bibinfo {pages} {196802} (\bibinfo {year} {2003})}\BibitemShut
	{NoStop}%
	\bibitem [{\citenamefont {Pioro-Ladrière}\ \emph {et~al.}(2008)\citenamefont
		{Pioro-Ladrière}, \citenamefont {Obata}, \citenamefont {Tokura},
		\citenamefont {Shin}, \citenamefont {Kubo}, \citenamefont {Yoshida},
		\citenamefont {Taniyama},\ and\ \citenamefont
		{Tarucha}}]{pioro-ladriereElectricallyDrivenSingleelectron2008}%
	\BibitemOpen
	\bibfield  {author} {\bibinfo {author} {\bibfnamefont {M.}~\bibnamefont
			{Pioro-Ladrière}}, \bibinfo {author} {\bibfnamefont {T.}~\bibnamefont
			{Obata}}, \bibinfo {author} {\bibfnamefont {Y.}~\bibnamefont {Tokura}},
		\bibinfo {author} {\bibfnamefont {Y.-S.}\ \bibnamefont {Shin}}, \bibinfo
		{author} {\bibfnamefont {T.}~\bibnamefont {Kubo}}, \bibinfo {author}
		{\bibfnamefont {K.}~\bibnamefont {Yoshida}}, \bibinfo {author} {\bibfnamefont
			{T.}~\bibnamefont {Taniyama}},\ and\ \bibinfo {author} {\bibfnamefont
			{S.}~\bibnamefont {Tarucha}},\ }\bibfield  {title} {{\selectlanguage
			{english}\bibinfo {title} {Electrically driven single-electron spin resonance in a
				slanting {Zeeman} field}},\ }\href {https://doi.org/10.1038/nphys1053}
	{\bibfield  {journal} {\bibinfo  {journal} {Nature Physics}\ }\textbf
		{\bibinfo {volume} {4}},\ \bibinfo {pages} {776} (\bibinfo {year}
		{2008})}\BibitemShut {NoStop}%
	\bibitem [{\citenamefont {Jardine}\ \emph {et~al.}(2021)\citenamefont
		{Jardine}, \citenamefont {Stenger}, \citenamefont {Jiang}, \citenamefont
		{de~Jong}, \citenamefont {Wang}, \citenamefont {Bleszynski~Jayich},\ and\
		\citenamefont {Frolov}}]{jardineIntegratingMicromagnetsHybrid2021}%
	\BibitemOpen
	\bibfield  {author} {\bibinfo {author} {\bibfnamefont {M.}~\bibnamefont
			{Jardine}}, \bibinfo {author} {\bibfnamefont {J.}~\bibnamefont {Stenger}},
		\bibinfo {author} {\bibfnamefont {Y.}~\bibnamefont {Jiang}}, \bibinfo
		{author} {\bibfnamefont {E.~J.}\ \bibnamefont {de~Jong}}, \bibinfo {author}
		{\bibfnamefont {W.}~\bibnamefont {Wang}}, \bibinfo {author} {\bibfnamefont
			{A.~C.}\ \bibnamefont {Bleszynski~Jayich}},\ and\ \bibinfo {author}
		{\bibfnamefont {S.~M.}\ \bibnamefont {Frolov}},\ }\bibfield  {title}
	{{\selectlanguage {english}\bibinfo {title} {Integrating micromagnets and hybrid
				nanowires for topological quantum computing}},\ }\href
	{https://doi.org/10.21468/SciPostPhys.11.5.090} {\bibfield  {journal}
		{\bibinfo  {journal} {SciPost Physics}\ }\textbf {\bibinfo {volume} {11}},\
		\bibinfo {pages} {090} (\bibinfo {year} {2021})}\BibitemShut {NoStop}%
	\bibitem [{\citenamefont {Zhang}\ \emph {et~al.}(2021)\citenamefont {Zhang},
		\citenamefont {Zhou}, \citenamefont {Hu}, \citenamefont {Ma}, \citenamefont
		{Ni}, \citenamefont {Wang}, \citenamefont {Luo}, \citenamefont {Cao},
		\citenamefont {Wang}, \citenamefont {Huang}, \citenamefont {Hu},
		\citenamefont {Jiang}, \citenamefont {Li}, \citenamefont {Guo},\ and\
		\citenamefont {Guo}}]{zhangControllingSyntheticSpinOrbit2021}%
	\BibitemOpen
	\bibfield  {author} {\bibinfo {author} {\bibfnamefont {X.}~\bibnamefont
			{Zhang}}, \bibinfo {author} {\bibfnamefont {Y.}~\bibnamefont {Zhou}},
		\bibinfo {author} {\bibfnamefont {R.-Z.}\ \bibnamefont {Hu}}, \bibinfo
		{author} {\bibfnamefont {R.-L.}\ \bibnamefont {Ma}}, \bibinfo {author}
		{\bibfnamefont {M.}~\bibnamefont {Ni}}, \bibinfo {author} {\bibfnamefont
			{K.}~\bibnamefont {Wang}}, \bibinfo {author} {\bibfnamefont {G.}~\bibnamefont
			{Luo}}, \bibinfo {author} {\bibfnamefont {G.}~\bibnamefont {Cao}}, \bibinfo
		{author} {\bibfnamefont {G.-L.}\ \bibnamefont {Wang}}, \bibinfo {author}
		{\bibfnamefont {P.}~\bibnamefont {Huang}}, \bibinfo {author} {\bibfnamefont
			{X.}~\bibnamefont {Hu}}, \bibinfo {author} {\bibfnamefont {H.-W.}\
			\bibnamefont {Jiang}}, \bibinfo {author} {\bibfnamefont {H.-O.}\ \bibnamefont
			{Li}}, \bibinfo {author} {\bibfnamefont {G.-C.}\ \bibnamefont {Guo}},\ and\
		\bibinfo {author} {\bibfnamefont {G.-P.}\ \bibnamefont {Guo}},\ }\bibfield
	{title} {\bibinfo {title} {Controlling {Synthetic} {Spin}-{Orbit} {Coupling}
			in a {Silicon} {Quantum} {Dot} with {Magnetic} {Field}},\ }\href
	{https://doi.org/10.1103/PhysRevApplied.15.044042} {\bibfield  {journal}
		{\bibinfo  {journal} {Physical Review Applied}\ }\textbf {\bibinfo {volume}
			{15}},\ \bibinfo {pages} {044042} (\bibinfo {year} {2021})}\BibitemShut
	{NoStop}%
	\bibitem [{\citenamefont {Nowack}\ \emph
		{et~al.}(2007{\natexlab{a}})\citenamefont {Nowack}, \citenamefont {Koppens},
		\citenamefont {Nazarov},\ and\ \citenamefont
		{Vandersypen}}]{nowackCoherentControlSingle2007}%
	\BibitemOpen
	\bibfield  {author} {\bibinfo {author} {\bibfnamefont {K.~C.}\ \bibnamefont
			{Nowack}}, \bibinfo {author} {\bibfnamefont {F.~H.~L.}\ \bibnamefont
			{Koppens}}, \bibinfo {author} {\bibfnamefont {Y.~V.}\ \bibnamefont
			{Nazarov}},\ and\ \bibinfo {author} {\bibfnamefont {L.~M.~K.}\ \bibnamefont
			{Vandersypen}},\ }\bibfield  {title} {\bibinfo {title} {Coherent {Control} of
			a {Single} {Electron} {Spin} with {Electric} {Fields}},\ }\href
	{https://doi.org/10.1126/science.1148092} {\bibfield  {journal} {\bibinfo
			{journal} {Science}\ }\textbf {\bibinfo {volume} {318}},\ \bibinfo {pages}
		{1430} (\bibinfo {year} {2007}{\natexlab{a}})}\BibitemShut {NoStop}%
	\bibitem [{\citenamefont {Nowack}\ \emph
		{et~al.}(2007{\natexlab{b}})\citenamefont {Nowack}, \citenamefont {Koppens},
		\citenamefont {Nazarov},\ and\ \citenamefont
		{Vandersypen}}]{nowack_coherent_2007}%
	\BibitemOpen
	\bibfield  {author} {\bibinfo {author} {\bibfnamefont {K.~C.}\ \bibnamefont
			{Nowack}}, \bibinfo {author} {\bibfnamefont {F.~H.~L.}\ \bibnamefont
			{Koppens}}, \bibinfo {author} {\bibfnamefont {Y.~V.}\ \bibnamefont
			{Nazarov}},\ and\ \bibinfo {author} {\bibfnamefont {L.~M.~K.}\ \bibnamefont
			{Vandersypen}},\ }\bibfield  {title} {\bibinfo {title} {Coherent control of a
			single electron spin with electric fields},\ }\href
	{https://doi.org/10.1126/science.1148092} {\bibfield  {journal} {\bibinfo
			{journal} {Science}\ }\textbf {\bibinfo {volume} {318}},\ \bibinfo {pages}
		{1430} (\bibinfo {year} {2007}{\natexlab{b}})},\ \bibinfo {note}
	{arXiv:0707.3080 [cond-mat]}\BibitemShut {NoStop}%
	\bibitem [{\citenamefont {Levitov}\ and\ \citenamefont
		{Rashba}(2002)}]{levitovDynamicalSpinelectricCoupling2002}%
	\BibitemOpen
	\bibfield  {author} {\bibinfo {author} {\bibfnamefont {L.~S.}\ \bibnamefont
			{Levitov}}\ and\ \bibinfo {author} {\bibfnamefont {E.~I.}\ \bibnamefont
			{Rashba}},\ }\href {https://doi.org/10.48550/arXiv.cond-mat/0209507}
	{\bibinfo {title} {Dynamical spin-electric coupling in a quantum dot}}
	(\bibinfo {year} {2002}),\ \bibinfo {note}
	{arXiv:cond-mat/0209507}\BibitemShut {NoStop}%
	\bibitem [{\citenamefont {Takase}\ \emph {et~al.}(2017)\citenamefont {Takase},
		\citenamefont {Ashikawa}, \citenamefont {Zhang}, \citenamefont {Tateno},\
		and\ \citenamefont {Sasaki}}]{takaseHighlyGatetuneableRashba2017}%
	\BibitemOpen
	\bibfield  {author} {\bibinfo {author} {\bibfnamefont {K.}~\bibnamefont
			{Takase}}, \bibinfo {author} {\bibfnamefont {Y.}~\bibnamefont {Ashikawa}},
		\bibinfo {author} {\bibfnamefont {G.}~\bibnamefont {Zhang}}, \bibinfo
		{author} {\bibfnamefont {K.}~\bibnamefont {Tateno}},\ and\ \bibinfo {author}
		{\bibfnamefont {S.}~\bibnamefont {Sasaki}},\ }\bibfield  {title}
	{{\selectlanguage {english}\bibinfo {title} {Highly gate-tuneable {Rashba}
				spin-orbit interaction in a gate-all-around {InAs} nanowire
				metal-oxide-semiconductor field-effect transistor}},\ }\href
	{https://doi.org/10.1038/s41598-017-01080-0} {\bibfield  {journal} {\bibinfo
			{journal} {Scientific Reports}\ }\textbf {\bibinfo {volume} {7}},\ \bibinfo
		{pages} {930} (\bibinfo {year} {2017})}\BibitemShut {NoStop}%
	\bibitem [{\citenamefont {Lewis}\ \emph {et~al.}(2022)\citenamefont {Lewis},
		\citenamefont {Banchi}, \citenamefont {Teoh}, \citenamefont {Islam},\ and\
		\citenamefont {Bose}}]{lewis_ion_2022}%
	\BibitemOpen
	\bibfield  {author} {\bibinfo {author} {\bibfnamefont {D.}~\bibnamefont
			{Lewis}}, \bibinfo {author} {\bibfnamefont {L.}~\bibnamefont {Banchi}},
		\bibinfo {author} {\bibfnamefont {Y.~H.}\ \bibnamefont {Teoh}}, \bibinfo
		{author} {\bibfnamefont {R.}~\bibnamefont {Islam}},\ and\ \bibinfo {author}
		{\bibfnamefont {S.}~\bibnamefont {Bose}},\ }\href
	{https://doi.org/10.48550/arXiv.2206.13685} {\bibinfo {title} {Ion {Trap}
			{Long}-{Range} {XY} {Model} for {Quantum} {State} {Transfer} and {Optimal}
			{Spatial} {Search}}} (\bibinfo {year} {2022}),\ \bibinfo {note}
	{arXiv:2206.13685 [quant-ph]}\BibitemShut {NoStop}%
	\bibitem [{\citenamefont {Waugh}\ and\ \citenamefont
		{Dolling}(1963)}]{PhysRev.132.2410}%
	\BibitemOpen
	\bibfield  {author} {\bibinfo {author} {\bibfnamefont {J.~L.~T.}\
			\bibnamefont {Waugh}}\ and\ \bibinfo {author} {\bibfnamefont
			{G.}~\bibnamefont {Dolling}},\ }\bibfield  {title} {\bibinfo {title} {Crystal
			dynamics of gallium arsenide},\ }\href
	{https://doi.org/10.1103/PhysRev.132.2410} {\bibfield  {journal} {\bibinfo
			{journal} {Phys. Rev.}\ }\textbf {\bibinfo {volume} {132}},\ \bibinfo {pages}
		{2410} (\bibinfo {year} {1963})}\BibitemShut {NoStop}%
	\bibitem [{\citenamefont {Dvir}\ \emph {et~al.}(2023)\citenamefont {Dvir},
		\citenamefont {Wang}, \citenamefont {van Loo}, \citenamefont {Liu},
		\citenamefont {Mazur}, \citenamefont {Bordin}, \citenamefont {ten Haaf},
		\citenamefont {Wang}, \citenamefont {van Driel}, \citenamefont {Zatelli},
		\citenamefont {Li}, \citenamefont {Malinowski}, \citenamefont {Gazibegovic},
		\citenamefont {Badawy}, \citenamefont {Bakkers}, \citenamefont {Wimmer},\
		and\ \citenamefont {Kouwenhoven}}]{dvirRealizationMinimalKitaev2023}%
	\BibitemOpen
	\bibfield  {author} {\bibinfo {author} {\bibfnamefont {T.}~\bibnamefont
			{Dvir}}, \bibinfo {author} {\bibfnamefont {G.}~\bibnamefont {Wang}}, \bibinfo
		{author} {\bibfnamefont {N.}~\bibnamefont {van Loo}}, \bibinfo {author}
		{\bibfnamefont {C.-X.}\ \bibnamefont {Liu}}, \bibinfo {author} {\bibfnamefont
			{G.~P.}\ \bibnamefont {Mazur}}, \bibinfo {author} {\bibfnamefont
			{A.}~\bibnamefont {Bordin}}, \bibinfo {author} {\bibfnamefont {S.~L.~D.}\
			\bibnamefont {ten Haaf}}, \bibinfo {author} {\bibfnamefont {J.-Y.}\
			\bibnamefont {Wang}}, \bibinfo {author} {\bibfnamefont {D.}~\bibnamefont {van
				Driel}}, \bibinfo {author} {\bibfnamefont {F.}~\bibnamefont {Zatelli}},
		\bibinfo {author} {\bibfnamefont {X.}~\bibnamefont {Li}}, \bibinfo {author}
		{\bibfnamefont {F.~K.}\ \bibnamefont {Malinowski}}, \bibinfo {author}
		{\bibfnamefont {S.}~\bibnamefont {Gazibegovic}}, \bibinfo {author}
		{\bibfnamefont {G.}~\bibnamefont {Badawy}}, \bibinfo {author} {\bibfnamefont
			{E.~P. A.~M.}\ \bibnamefont {Bakkers}}, \bibinfo {author} {\bibfnamefont
			{M.}~\bibnamefont {Wimmer}},\ and\ \bibinfo {author} {\bibfnamefont {L.~P.}\
			\bibnamefont {Kouwenhoven}},\ }\bibfield  {title} {{\selectlanguage
			{english}\bibinfo {title} {Realization of a minimal {Kitaev} chain in coupled
				quantum dots}},\ }\href {https://doi.org/10.1038/s41586-022-05585-1}
	{\bibfield  {journal} {\bibinfo  {journal} {Nature}\ }\textbf {\bibinfo
			{volume} {614}},\ \bibinfo {pages} {445} (\bibinfo {year}
		{2023})}\BibitemShut {NoStop}%
	\bibitem [{\citenamefont {Huang}\ and\ \citenamefont
		{Hu}(2021)}]{huang_spin_2021}%
	\BibitemOpen
	\bibfield  {author} {\bibinfo {author} {\bibfnamefont {P.}~\bibnamefont
			{Huang}}\ and\ \bibinfo {author} {\bibfnamefont {X.}~\bibnamefont {Hu}},\
	}\bibfield  {title} {{\selectlanguage {english}\bibinfo {title} {Spin manipulation
				and decoherence in a quantum dot mediated by a synthetic spin–orbit
				coupling of broken {T}-symmetry}},\ }\href
	{https://doi.org/10.1088/1367-2630/ac430c} {\bibfield  {journal} {\bibinfo
			{journal} {New Journal of Physics}\ }\textbf {\bibinfo {volume} {24}},\
		\bibinfo {pages} {013002} (\bibinfo {year} {2021})},\ \bibinfo {note}
	{publisher: IOP Publishing}\BibitemShut {NoStop}%
\end{thebibliography}
\end{document}